\documentclass[aps,pra,10pt,notitlepage,nofootinbib,tightenlines,twocolumn,superscriptaddress]{revtex4-2}

\usepackage{bm,bbm,amssymb,amsmath,amsfonts,amsthm,mathrsfs,MnSymbol, times}
\usepackage{natbib}
\usepackage{graphicx}
\usepackage{amsmath}
\usepackage{multirow}

\usepackage{color}
\usepackage{bbold}
\usepackage[utf8]{inputenc}
\usepackage{empheq}
\usepackage{soul}
\usepackage[ampersand]{easylist} 
\usepackage[normalem]{ulem}
\usepackage{braket}
\usepackage{hyperref}
\hypersetup{
    colorlinks=true,       
    linkcolor=red,          
    citecolor=magenta,        
    filecolor=magenta,      
    urlcolor=cyan,           
    runcolor=cyan
}
\graphicspath{ {figures/} }

\newcommand{\eq}[1]{\begin{align}#1\end{align}}

\newcommand{\eqs}[1]{\begin{align}\begin{split}#1\end{split}\end{align}}

\newcommand{\bel}{\begin{easylist}[itemize]}
\newcommand{\eel}{\end{easylist}}
\newcommand{\vsigma}{\vec{\sigma}}
\newcommand{\sx}{\sigma_x}
\newcommand{\sy}{\sigma_y}
\newcommand{\sz}{\sigma_z}
\newcommand{\vbeta}{\vec{\beta}}
\newcommand{\vOmega}{\vec{\Omega}}
\newcommand{\hn}{\hat{n}}
\newcommand{\pr}[1]{\left(#1\right)}

\newcommand{\mrm}{\mathrm}
\newcommand{\mB}{\mathcal{B}}
\newcommand{\mF}{\mathcal{F}}
\newcommand{\CNB}{\overline{\mathcal{B}}}
\newcommand{\id}{\mathbb{1}}
\newcommand{\tU}{\tilde{U}}
\newcommand{\tr}{\mathrm{Tr}}

\newcommand{\RR}{\mathcal{R}}
\newcommand{\tH}{\tilde{H}}

\DeclareMathOperator{\sinc}{sinc}

\definecolor{ao(english)}{rgb}{0.0, 0.5, 0.0}
\newcommand{\red}[1]{{\color{red} {#1}}}

\begin{document}
\title{Optimally Band-Limited Noise Filtering for Single Qubit Gates}
\author{Yasuo Oda}
\affiliation{William H. Miller III Department of
Physics $\&$ Astronomy, Johns Hopkins University, Baltimore, Maryland 21218, USA}
\author{Dennis Lucarelli}
\affiliation{Department of Physics, American University}
\author{Kevin Schultz}
\affiliation{Johns Hopkins University Applied Physics Laboratory}
\author{B. David Clader}\thanks{Current affiliation: Goldman Sachs \& Co.}
\affiliation{Johns Hopkins University Applied Physics Laboratory}
\author{Gregory Quiroz}
\affiliation{Johns Hopkins University Applied Physics Laboratory}
\affiliation{William H. Miller III Department of
Physics $\&$ Astronomy, Johns Hopkins University, Baltimore, Maryland 21218, USA}

\begin{abstract}
    We introduce a quantum control protocol that produces smooth, experimentally implementable control sequences optimized to combat temporally correlated noise for single qubit systems.
    The control ansatz is specifically chosen to be a functional expansion of discrete prolate spheroidal sequences, a discrete time basis known to be optimally concentrated in time and frequency, and quite attractive when faced with experimental control hardware constraints. 
    We leverage the filter function formalism to transform the control problem into a filter design problem, and show that the frequency response of a quantum system can be carefully tailored to avoid the most relevant dynamical contributions of noise processes. 
    Using gradient ascent, we obtain optimized filter functions and exploit them to elucidate important details about the relationship between filter function design, control bandwidth, and noise characteristics. In particular, we identify regimes of optimal noise suppression and in turn, optimal control bandwidth directly proportional to the size of the frequency bands where the noise power is large. In addition to providing guiding principles for filter design, our approach enables the development of controls that simultaneously yield robust noise filtering and high fidelity single qubit logic operations in a wide variety of complex noise environments.
\end{abstract}
\maketitle

\section{Introduction}

The ability to perform fast and robust operations on multi-qubit quantum systems is a necessity for realizing reliable quantum computation~\cite{Preskill2018quantumcomputingin}.
Unfortunately, the inevitable interaction between a quantum system and its environment presents an obstacle for achieving such operations. Unwanted system-environment interactions lead to noise processes that cause quantum gates to deviate from their intended evolution, consequently leading to a loss of coherence and computational errors. 
Quantum control is an approach that seeks address this challenge through the design of control protocols that implement desired quantum operations, while simultaneously achieving robustness against noise~\cite{dong2010quantum}. Quantum control can be particularly advantageous for combating spatio-temporally correlated noise, which is known to be detrimental to quantum error correction~\cite{Clemens2004qec,Terhal2005qec,Aharonov2006qec, Alicki2006qec,Ng2009qec}.

Various control techniques have been developed to carry out robust quantum gates in the presence of systematic and environmental noise sources. Pulse-based techniques such as dynamically corrected gates~\cite{Khodjasteh2009DCGs,Khodjasteh2012DCGs} leverage features of dynamical decoupling~\cite{Viola1999DD,lidar2013book} to effectively average out noise while implementing a logical operation. Despite their ability to account for practical limitations, such as bounded control amplitudes, they are limited to static noise models. Smooth control methods based on quantum optimal control theory, such as open-system Gradient Ascent Pulse Engineering (GRAPE)~\cite{abdelhafez:2019-grape}, extend beyond the traditional closed system GRAPE approaches~\cite{skinner2003grape,Khaneja2005GRAPE,wilhelm2020introduction} to enable the construction of quantum gates in the presence of time-dependent noise. Open-system GRAPE performs local updates to the control waveform in accordance with typical GRAPE approaches, however, requires averaging over dynamical simulations of quantum trajectories to optimize control waveforms in the time domain.

The \emph{filter function formalism} (FFF) offers an alternative perspective on optimized quantum control in the presence of time-dependent noise processes. Capable of accommodating a wide range of spatio-temporally correlated noise models, the FFF captures a quantum system's sensitivity to noise in the frequency domain via control-dependent filter functions (FFs)~\cite{Cywinski2008ff,Green2013ff,PazSilva2014ff}. Within the FFF, gate fidelity can be expressed in terms of an overlap integral between the FFs and the noise power spectral densities (PSDs). This relationship gives rise to a highly intuitive perspective on robust quantum gates, namely, minimizing spectral overlap between the FFs and noise PSDs is essential for realizing noise-optimized gates. 

Minimization of the overlap integral has become a guiding principle for gate optimization via FF design. Proposed approaches have made direct use of the overlap integral as an objective function~\cite{Ball2015Walsh, Le2022FilterOpt}, where a priori knowledge of the noise PSD is assumed, or focused on the minimization of the FF over a specified low-frequency band~\cite{Ball2015Walsh,Huang2017RobustQuantumGates}. In practice,  both objective functions require estimation of the noise PSD (e.g., through quantum noise spectroscopy (QNS)~\cite{Alvarez2011qns, Paz-Silva2017MutilqubitSpectroscopy, Norris2018qns}), with the latter potentially requiring only knowledge of the noise cutoff frequencies. Regardless of the choice of objective function, a majority of the approaches have been numerically-oriented, utilizing either gradient~\cite{Le2022FilterOpt} or non-gradient-based optimization~\cite{Ball2015Walsh, Huang2017RobustQuantumGates}. While analytical solutions for optimal FFs are difficult to ascertain due to the non-linear relationship between the control and FF, numerical approaches have offered little intuition into the design of optimal FFs. Furthermore, questions remain regarding the interplay between control parameters, such as bandwidth and amplitude, noise parameters (e.g., noise cutoff frequencies), and optimal FF design.

\begin{figure*}[t]
\centering
\includegraphics[width=\textwidth,trim={0 0 0 0cm},clip]{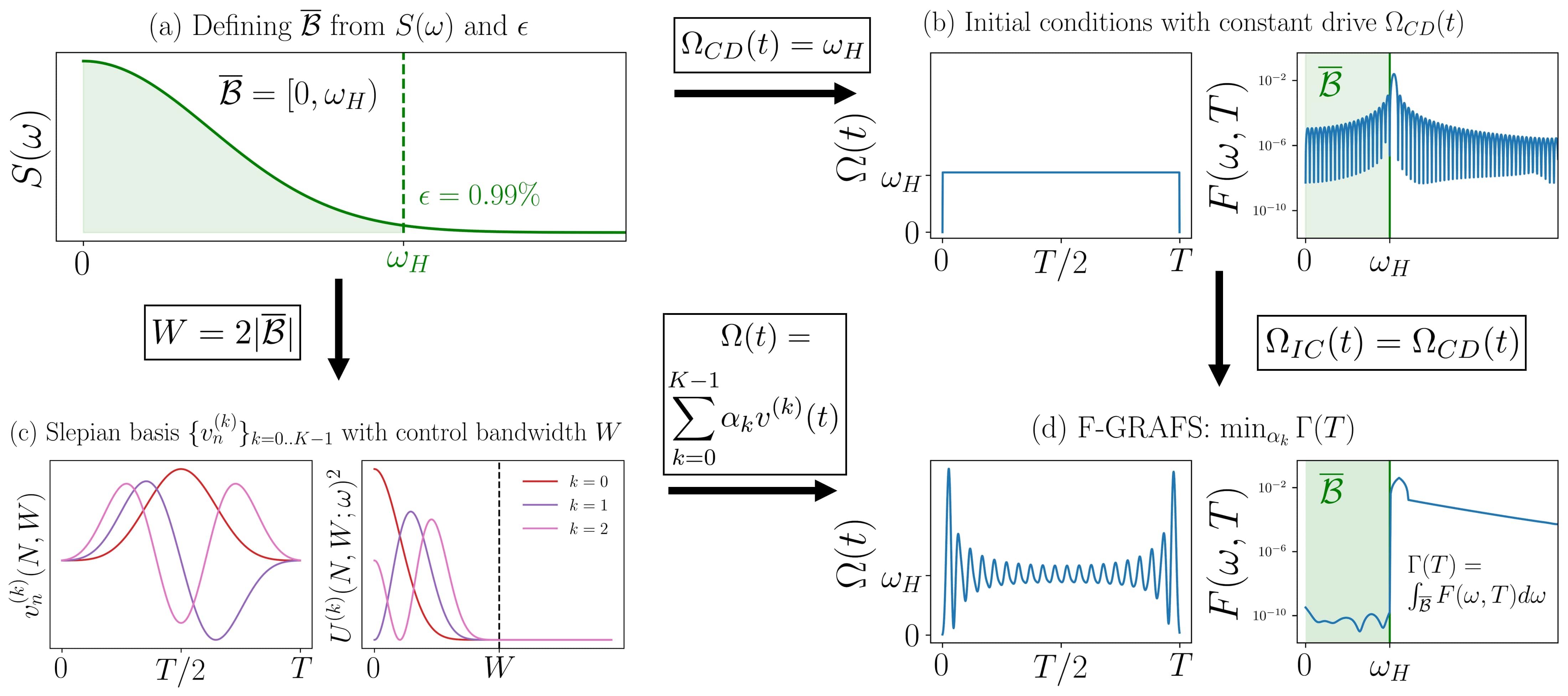}
\caption{
F-GRAFS workflow, exemplified for single axis noise and control case. (a) A CNB is defined from the PSD $S(\omega)$ and a fractional power $\epsilon$, as the region where the noise is the strongest. (b) Using the maximum frequency $\omega_H=\max_\omega \CNB$, the initial conditions are defined as constant drive. (c) The size of the CNB is used to define the control bandwidth of the Slepian basis. (d) The initial condition is projected onto the Slepian basis, and optimized to minimize the objective function $\Gamma(T)$, improving the cancelation of the noise by several orders of magnitude. }
\label{fig:intro-fig}
\end{figure*}

In this work, we provide analytical insight into FF design and introduce an optimization protocol that sheds light on the relationship between control, noise, and optimal FFs. With focus on a single qubit system subject to additive dephasing, we show that relatively simple control schemes can provide FF tunability in both single and multi-axis noise scenarios. Moreover, we show that such schemes can be straightforwardly designed based on the properties of the noise PSD. Analytically designed controls are used to inform the initialization of a FF optimization approach we refer to as Filter GRadient Ascent in Function Space (F-GRAFS). As an extension of the GRAFS method~\cite{2018Lucarelli_GRAFS}, F-GRAFS seeks to minimize the spectral support of the FF within a specified frequency band while simultaneously performing a non-trivial quantum gate. F-GRAFS is shown to be highly versatile and adaptable to a variety of multi-axis noise scenarios, including non-uniform high-pass and band-pass gates. Furthermore, it proves to be a valuable tool for examining the dependence of FF design on control and noise parameters.

Following the GRAFS approach, we utilize the discrete prolate spheroidal sequences (DPSS) or so-called ``Slepians" ~\cite{slepian1961prolate,landau1961prolate,landau1962prolate,slepian1964prolate,papoulis1972digital, slepian1978prolate} as a functional basis for expressing the control waveform. The DPSS possess intrinsic bandwidth tunability that enables the study of optimized FFs as a function of control bandwidth. Additionally, the DPSS bases constitute an optimal description of the subspace of functions limited in both time and bandwidth. Using the DPSS as functional basis restricts the controls to the space of physically realizable functions, while substantially reducing the dimensionality of the optimization problem. It is through the use of F-GRAFS in conjunction with the DPSS that we arrive at conditions on optimal control bandwidth for filter design. In particular, we find clear indications that the optimal control bandwidth is lower-bounded by twice the size of the frequency band over which the FF is to be suppressed. Interestingly, this result generally holds for both single and multi-axis noise. 

Together, our analytical and numerical results provide a relatively comprehensive guide for FF design in a variety of relevant noise scenarios.
Fig. \ref{fig:intro-fig} summarizes the F-GRAFS workflow. First, the region of frequencies where the noise is strongest is identified. Then, this information is used to construct the DPSS basis and tailor initial conditions, both key ingredients in the gradient-based optimization.

The manuscript is organized as follows. 
In Sec.~\ref{sec:theory} we describe the relevant background necessary to understand the numerical optimization, namely the system models and the FFF. 
We also describe the DPSS sequences, their definition, and main properties.
In Sec.~\ref{sec:F-GRAFS}, we introduce the F-GRAFS method, defining the optimization problem and explicitly computing the gradients.
Sec.~\ref{sec:results} presents the main results. 
Here, we describe analytical control schemes used to initial F-GRAFS. We then showcase optimized control waveforms and FFs obtained from F-GRAFS for various control and noise scenarios. F-GRAFS is then employed to examine the connection between control parameters, noise characteristics, and optimal FF design.
We conclude in Sec.~\ref{sec:conclusion} with a summary of the main results and an outline of future investigations.

\section{Background}
\label{sec:theory}
We begin by describing the theory necessary to understand and implement F-GRAFS.
First, we define the control and noise Hamiltonians relevant to this study.
Then, we describe the FFF, the framework used to study the system dynamics in the frequency domain.

\subsection{System Model}
We consider the problem of controlling a single qubit system in the presence of temporally correlated additive dephasing noise. The dynamics of the system are governed by a Hamiltonian $H(t)$ that can be partitioned as 
\eq{
\label{eq:H_sys}
H(t) = H_C(t)+H_N(t),
}
where $H_C(t)$ denotes the control contribution that acts solely on the system and $H_N(t)$ encapsulates the contributions of the noise processes.
In the reference frame rotating with the frequency of the qubit, and under the rotating wave approximation, the control Hamiltonian is given by (in units of $\hbar=1$)
\eq{
H_C(t)=\frac{\Omega_x(t)}{2}\sx + \frac{\Omega_y(t)}{2}\sy.
\label{eq:H_Ctrl}
}
Motivated by a variety of system architectures~\cite{Souza2012dd, soare2014experimental, krantz2019quantum}, we consider control functions along the axes transverse to the quantization direction.
While the control Hamiltonian defines amplitude control along the Pauli operators $\sx$ and $\sy$, one can move from Cartesian to polar coordinates to express the control functions in terms of time-dependent amplitude and phase. 
We focus on the former representation, yet note that our approach is effectively agnostic to the choice in control representation. Formally, the controls are assumed to be expressed as a weighted sum
\begin{equation}
    \Omega_\nu(t)=\sum_k \alpha_{\nu, k}\, \varphi_k(t).
    \label{eq:ctrl-param}
\end{equation}
The expansion coefficients $\alpha_{\nu, k}$ are real and weight basis functions $\varphi_k(t)$. At this stage, we assume $\varphi_k(t)$ to be arbitrary and more concretely specify them in Sec.~\ref{subsec:dpss}. 

The single qubit system is subject to semiclassical noise generically described by the Hamiltonian
\eq{
H_N(t)=\vbeta(t)\cdot\vsigma,
\label{eq:H-noise}
} 
where $\vbeta(t) = (\beta_{x}(t),\beta_y(t),\beta_z(t))$ and the Pauli operators are given by $\vec{\sigma}=(\sx, \sy, \sz)$.
Each noise component $\beta_\mu(t)$ defines a random Gaussian process considered to be wide sense stationary with zero mean, $\braket{\beta_\mu(t)}=0, \mu=x,y,z$, where $\braket{\cdots}$ denotes classical ensemble averaging. 
In addition, the functions $\beta_\mu(t)$ are characterized by two-point correlation functions $\langle \beta_\mu(t)\beta_{\nu}(t') \rangle$, related to the noise PSD $S_{\mu\nu}(\omega)$ via a Fourier transform:
\eq{
\label{eq:2pt_PSD_general}
\langle \beta_\mu(t)\beta_{\nu}(t') \rangle = \frac{1}{2\pi}\int_{0}^{\infty} S_{\mu\nu}(\omega) e^{i\omega (t-t')} d\omega.
}
As will be discussed below, the frequency domain representation provides a convenient language for analyzing the dynamical contributions of the noise to the time evolution of the qubit.
This representation gives rise to a powerful framework known as the FFF. 
Capturing the effective dynamics of a quantum system in terms of the spectral properties of the noise and the control-driven frequency response of the system, the FFF has been employed in settings such as quantum noise spectroscopy~\cite{Alvarez2011qns, Paz-Silva2017MutilqubitSpectroscopy, Norris2018qns} and noise mitigation~\cite{2011Ajoy_alvarezDD,Biercuk2011ff,Ball2015Walsh,soare2014experimental}.
In the following section, we will review this formalism, following closely the work by \emph{Green et al.} \cite{2013Green}.

\subsection{Filter Function Formalism}
\label{sec:FFF}
Due to the time dependent nature of the noise, the Hamiltonian in Eq.~(\ref{eq:H_sys}) will in general not commute with itself at different times $[H(t),H(t')]\neq0$ if $t\neq t'$, and
therefore the time evolution it induces will be given by the time-ordered propagator 
\eq{
\label{eq:time_propagator}
U(t) = \mathcal{T}_+ \exp\left[ -i\int_0^t H(s) ds \right].
}
where $\mathcal{T}_+$ is the time-ordering operator.
In general, for an arbitrary Hamiltonian $H(t)$, the time propagator $U(t)$ will not be analytically tractable and hence we will not have access to a closed analytical description of the time evolution. 
Instead, by moving to the interaction picture with respect to the control, and assuming that the control is dominant with respect to the noise, the noise dynamics can be treated as a time-dependent perturbation.

Moving into the rotating reference frame with respect to the control propagator $U_C(t) = \mathcal{T}_+ \exp\left( -i\int_0^t H_C(s) ds \right)$, 
the time evolution operator $U(t)$ can be expressed as $U(t)=U_C(t)\tilde{U}_N(t)$. 
The rotated-frame error propagator $\tilde{U}_N(t) = \mathcal{T}_+ \exp\left( -i\int_0^t \tilde{H}_N(s) ds \right)$ is generated by 
\begin{eqnarray}
\tilde{H}_N(t)=U_C^\dagger(t)H_N(t)U_C(t)
=\sum_{\mu\nu} \beta_\mu(t)  R_{\mu\nu}(t) \sigma_\nu.
\end{eqnarray}
Since $SU(2)$ is homomorphic to $SO(3)$, the rotated-frame Hamiltonian can be written in terms of the \emph{control matrix} components $R_{\mu\nu}(t)$. Each component can be expressed using the Hilbert-Schmidt inner product as
\eq{
\label{eq:control_matrix}
R_{\mu\nu}(t)=\frac{1}{2}\mathrm{Tr}[U_C^\dagger(t)\sigma_\mu U_C(t)\sigma_\nu].
}

Using a perturbative Magnus expansion~\cite{blanes2009magnus}, it is convenient to parametrize the error propagator
\begin{equation}
    \tilde{U}_N(t)=\exp[{-i \vec{a}(t)\cdot \vec{\sigma}}]
\end{equation}
in terms of the \emph{error vector} $\vec{a}(t)=\sum^{\infty}_{l=1}\vec{a}^{(l)}(t)$. In general, a closed form for $\vec{a}(t)$ does not exist, however, when the noise is sufficiently weak and the time scale of the dynamics is sufficiently short, one can truncate the expansion~\cite{Green2013ff,Norris2018qns}. Under these conditions, the error vector can be approximated to leading (first) order such that $\vec{a}(t)\approx \vec{a}^{(1)}(t)$, where
\begin{equation}
a^{(1)}_\nu(t) = \sum_{\mu} \int_0^t ds\, \beta_\mu(s) R_{\mu\nu}(s),
    \label{eq:error_vector}
\end{equation}
with $\mu,\nu=x,y,z$. 

The error vector representation proves to be convenient for examining the efficacy of a control protocol via the average operational fidelity
\begin{equation}
\mathcal{F}(T)\equiv \frac{1}{4}\braket{| \mathrm{Tr} (U^\dagger_\mathrm{G} U(T)) |^2 }.
\label{eq:op-fidelity}
\end{equation}
This particular measure utilizes the Hilbert-Schmidt inner product to quantify how well a given (noisy) propagator $U(T)$ approximates a target gate $U_G$ after a total controlled evolution time $T$. In the case where $U_C(T)=U_G$, this measure can be expressed as
\begin{equation}
    \mathcal{F}_{N}(T) = \frac{1}{2}\left( 1+e^{-\chi(T)} \right),
    \label{eq:fidelity_overlap}
\end{equation}
where $\chi(T)=\braket{|\vec{a}(T)|^2}$~\cite{Green2013ff}. Typically, $\chi(T)$ is referred to as the \emph{overlap} due to its frequency domain representation. More specifically, $\chi(T)$ can be generically expressed as a sum of products of integrals, where each integral is defined as a product between noise PSDs and the filter functions (FFs). In the weak noise limit, $\chi(T)\approx \braket{|\vec{a}^{(1)}(T)|^2}$ and the overlap conveniently reduces to 
\begin{equation}
\chi(T)\approx\frac{1}{\pi}\sum_{\mu=x,y,z} \int_{0}^{\infty} S_{\mu}(\omega) F_{\mu}(\omega,T)d\omega.
\label{eq:overlap_general}
\end{equation} 
The FFs are defined as 
\begin{equation}
    F_{\mu}(\omega,T)=\sum_{\nu} |R_{\mu\nu}(\omega,T)|^2,
    \label{eq:FF_R_general}
\end{equation}
where the frequency domain control matrix elements are
\begin{equation}
    R_{\mu\nu}(\omega,T)=\int_0^T R_{\mu\nu}(t)e^{i\omega t}dt.
    \label{eq:control_matrix_freq}
\end{equation}
Note that in this formulation, it is assumed that cross-correlations are neglected; thus, $S_\mu(\omega)=S_{\mu\nu}(\omega)\delta_{\mu\nu}$.

The FFF offers an alternative perspective that can be exploited for characterization and control problems. For example, in the case of optimized control, the objective is to find control functions $\Omega_\nu(t)$ that minimize the overlap [Eq.~(\ref{eq:overlap_general})], and thus, maximize the operational fidelity [Eq.~(\ref{eq:fidelity_overlap})]. This is the overarching principle leveraged by F-GRAFS to tailor FFs and achieve optimized gate. 

\subsection{Time-Band-Limited Sequences for Quantum Control}
\label{subsec:dpss}
Generally speaking, there is no designated protocol for choosing a parametrization of the control function $\Omega_\nu(t)$. In GRAPE approaches, the control profiles are typically assumed to be piecewise constant in time~\cite{skinner2003grape}. The optimization proceeds by locally updating each control amplitude for each timestep such that the overall profile generates a controlled evolution that converges towards the desired operation. This approach becomes increasingly computationally expensive as the number of timesteps increases. Furthermore, GRAPE methods typically require additional bandwidth and amplitude constraints to enforce physical limitations in control hardware or low-pass filtering to generate smooth control~\cite{wilhelm2020introduction}.

Functional expansions of the control waveform offer advantages over piecewise control. When expressed as a weighted sum of basis functions, control optimization algorithms focus their attention on optimizing the basis function weights, rather than the individual control amplitudes. This alternative approach leads to global, as opposed to local updates to the waveform. While GRAPE based methods~\cite{Machnes2018Tunable} and other non-gradient based methods~\cite{Huang2017RobustQuantumGates} have sought to leverage functional expansions for optimized control, basis selection is to some degree unmotivated. 

In this study, we employ a functional expansion parametrized by the DPSS~\cite{slepian1978prolate}. With their rich history in classical signal processing, DPSS offer an optimal compromise to the time-bandwidth uncertainty relation. More specifically, they form a basis with optimal spectral concentration for time-limited signals. From a control perspective, DPSS are attractive for designing optimized control that inherently account for physical limitations of control hardware. Timing resolution and control bandwidth preclude the basis generation. As a result, intrinsic bandwidth constraints are imposed within the basis prior to optimization rather than as an additional constraint appended to the objective function. 

As discrete analogs of prolate spheroidal wave functions, DPSS are parametrized by the sequence length $N$ and the dimensionless bandwidth parameter $W\in(0,0.5)$. A $k$th order Slepian sequence $\{v_m^{(k)}(N,W)\}_{m=0}^{N-1}$ is generated as a solution to the Toeplize matrix eigenvalue equation
\begin{equation}
\sum_{m=0}^{N-1} \frac{\sin2\pi W(n-m)}{\pi(n-m)} v_m^{(k)}(N,W) = \lambda_k(N,W) v_n^{(k)}(N,W),
\label{eq:slepian_eigen_eq}
\end{equation}
where $k,n=0,...,N-1$. The DPSS form an orthonormal basis of the vector space of real numbers $\mathbb{R}^N$, satisfying $\sum^{N-1}_{n=0} v^{(k)}_n(N,W)v^{(l)}_n(N,W)=\delta_{k,l}$. The eigenvalues $\{\lambda_k(N,W)\}$ determine the order $k$ of the DPSS and increase monotonically with $k$, such that $0\leq \lambda_0(N,W) \leq \lambda_1(N,W)\leq \cdots\leq \lambda_{N-1}(N,W)$. Moreover, $\{\lambda_k(N,W)\}$ are a measure of spectral concentration of DPSS within the frequency band $(-2\pi W/\delta t,2\pi W/\delta t)$, where $\delta t=T/N$ designates the time resolution of the control. DPSS of order $k < 2NW$ are the most spectrally concentrated, possessing eigenvalues very close to unity. In contrast, DPSS with $k \geq 2NW$ are characterized by $\lambda_k(N,W)\approx 0$. This property has been previously used to establish an approximate dimension $K$ of the space of band-limited functions: $K=\lfloor 2NW \rfloor$. Lastly, we note that the order $k$ of the DPSS determines its number of zero-crossings and  characterize even-odd symmetry of the sequence about the midpoint.

Below, the DPSS are used to parametrize the space of control functions available to the optimization algorithm. The size of the basis is dictated by $K$ and therefore, the timing resolution and bandwidth parameter. However, as we will discuss, leveraging the spectral information contained within the DPSS eigenvalues, we can considered a basis smaller than $K$ to introduce additional control constraints, e.g., endpoint constraints. We further exploit the DPSS to establish a relationship between the bandwidth $W$ and the noise suppression characteristics of our optimized gates.

\section{Filter GRAFS}
\label{sec:F-GRAFS}
F-GRAFS is a gradient-based optimization method for constructing noise-robust quantum operations via the FFF. Inspired by our previous work on closed system optimized control~\cite{2018Lucarelli_GRAFS}, F-GRAFS utilizes a functional expansion of the control in terms of DPSS. Below, we further elaborate on F-GRAFS, providing detailed information about the objective function, gradient expressions, and the optimization procedure.

\subsection{Optimization Problem}
\label{subsec:opt-prob}
F-GRAFS is designed to engineer noise-optimized control profiles that minimize the distance between a target gate $U_G$ and a noisy controlled evolution described by the unitary $U(T)=U_C(T)\tilde{U}(T)$. This is accomplished by casting the optimized control problem as a constrained optimization problem. The objective function aims to minimize the spectral overlap between the noise PSDs and FFs, while the constraint works to enforce a targeted fidelity for the logic gate. Formally, the F-GRAFS optimization problem is defined as
\begin{equation}
\begin{aligned}
    \min_{\{\alpha_k\}} \quad& \Gamma(T)\\
    \textrm{s.t.} \quad& \mathcal{F}_G(T)\geq 1-\epsilon_G
\end{aligned}.
\label{eq:opt-problem}
\end{equation}
where $\Gamma$ quantifies the \emph{spectral leakage} of the FFs within the frequency bands $\overline{\mathcal{B}}_{j}$. The constraint is defined relative to the ideal gate fidelity $\mathcal{F}_G(T)=\frac{1}{4}|\tr[U^\dagger_G U_C(T)]|^2$, which determines how well $U_C(T)$ approximates the desired target gate within an infidelity tolerance $\epsilon_G$.

The spectral leakage is dictated by the spectral null-bands (NBs) of the noise PSDs and the FFs resulting from candidate control profiles. The NBs are defined as the regions $\mathcal{B}_\mu$, $\mu=x,y,z$, where the fractional noise power is small. More rigorously, for a desired fractional noise power $\epsilon_\mu\ll 1$, the NB is defined according to
\eq{
\label{eq:fractional-power}
\frac{\int_{\mathcal{B}_{\mu}} S_{\mu}(\omega) d\omega}{\int_{0}^{\infty} S_{\mu}(\omega) d\omega} < \epsilon_\mu.
}
It is sufficient to consider equivalent fractional powers across all Pauli channels, and therefore, $\epsilon_\mu=\epsilon$ will be assumed throughout the remainder of this study. 

In general, the NB is selected based on three guidelines. (1) The NB should achieve maximum connectedness, i.e., minimize the number $L$ of disjoint sets that compose the NB: $\mB=\uplus_{\mu=0}^L \mB_\mu$. (2) The size of the NB $|\mB|$ should be maximal. As we will discuss later in this study, the control bandwidth requirements decrease with $|\mB|$. (3) The NB should be chosen to maximize the presence of high-frequency contributions. In practice, most time-correlated noise processes are characterized by PSDs that are concentrated at low frequency. A choice of $\epsilon$ will then determine a high-frequency cutoff $\omega_H$, producing $\mB=[\omega_H,\pi/\delta t)$, within which the noise is sufficiently weak.

It is within the NBs that the FFs would ideally reside. Thus, an optimized control scheme strives to maximize the FF support within the NB, or equivalently, minimize the leakage of the FFs within the complement of the NBs (CNBs) $\overline{\mathcal{B}}_{\mu}$. The F-GRAFS approach operates within the context of the latter, seeking to minimize the spectral leakage
\eq{
\Gamma(T) = \frac{1}{3T}\sum_{\mu=x,y,z}\overline{p}_\mu\int_{\overline{\mathcal{B}}_{\mu}} F_{\mu}(\omega,T) d\omega.
\label{eq:objective_function}
}
Here we have introduced the estimated weight of the noise power in the $\mu$-th direction $\overline{p}_\mu$; see Appendix~\ref{sec:appendix_objective_function} for further details. Observe that we have included a factor of $1/3T$ to normalize $\Gamma(T)$ with respect to the total power of the FFs. Each component FF possess a total power of $T$, while the factor of three appears due to presence of noise along all three single qubit Pauli channels.

While our focus will be on the spectral leakage, we show that there is connection between the minimization of $\Gamma(T)$ and the size of the CNB. Formally, we define the size of the CNB $|\overline{\mathcal{B}}|$ as the integral
\eq{
\label{eq:size_CNB}
|\overline{\mathcal{B}}| = \sum_{\mu=x,y,z} \int_{\overline{\mathcal{B}}_\mu}d\omega
} 
Each integral is bounded between 0 and $\pi/\delta t$ and thus, $0\leq |\overline{\mathcal{B}}| \leq 3\pi/\delta t$. The lower bound is saturated in the noiseless case, while the upper bound is achieved in the white noise case. In Sec.~\ref{sec:results}, the size of the CNB will emerge as an important quantity in the discussion of optimized spectral leakage and optimal control bandwidth.

The F-GRAFS optimization problem is solved via Sequential Least SQuares Programming (SLSQP)~\cite{kraft1988slsqp}. In practice, we find that SLSQP offers faster convergence rates than alternative numerical optimizers, such as the interior point method~\cite{potra2000ip} when solving Eq.~(\ref{eq:opt-problem}). Variants of this optimization problem, for example, utilizing an effective ``leakage fidelity" $\mathcal{F}_\Gamma(T)=\frac{1}{2}[1+\exp({-\frac{PT}{\delta\omega}\Gamma(T))}]$, where $P=\sum_\mu \int_{0}^{\infty} S_\mu(\omega)d\omega $ is the total power, leads to improved convergence for both the interior point method and L-BFGS-B~\cite{zhu1997lbfgsb}. However, the latter approach requires knowledge of the total power and therefore more detailed estimates of the noise PSDs. This is in contrast to the spectral leakage, which may only require rough estimates of noise PSDs to determine CNBs. For this reason, and its simplicity, we utilize Eq.~(\ref{eq:opt-problem}) for filter design and gate optimization.

While presented in a rather axiomatic fashion, the optimization problem given in Eq.~(\ref{eq:opt-problem}) can be shown to be related to the global phase invariant metric between unitaries~\cite{Fowler2011ConstructingAS,Kliuchnikov2016PracticalAO,Gheorghiu2021TcountAT,Mukhopadhyay2021ComposabilityOG}. This metric is defined as
\begin{eqnarray}
\label{eq:op-dist}
    D(U_G,U(T)) &=& \min_{\alpha\in[0,2\pi]} \|U_G e^{i\alpha}-U(T)\|_2 \nonumber\\
    &=& \sqrt{1-\frac{1}{2}\left| \tr{\left[U_G^\dagger U(T)\right]} \right|},
\end{eqnarray}
where $||A||_2 = \sqrt{\tr(A^\dagger A)}$ is the Frobenius norm. As a distance metric, $D$ naturally satisfies the properties of symmetry and the identity of indiscernibles. In addition, $D$ satisfies the triangle inequality, which can be used to establish the following upper bound on the average squared-distance:
\begin{eqnarray}
\braket{D(U_G,U(T))^2} &\leq& \braket{\left( D(U_G,U_C(T)) + D(\id ,\tU(T)) \right)^2} \nonumber\\
&\leq& 2-\mathcal{F}_G(T) - \mathcal{F}_\Gamma(T) + \mathcal{K}(T) + \mathcal{O}(\epsilon). \nonumber\\
\label{eq:dist-bound}
\end{eqnarray}
The term $\mathcal{K}(T)$ is a function of $\mF_G, \mF_N$ and goes to zero as these quantities approach unity. The last term signifies a dependence on the fractional power in the NB.
Note that as long as $\epsilon$ can be kept sufficiently small, the bound on $D$ is effectively minimized by minimizing $\Gamma(T)$ subject to a $\epsilon_G\ll 1$. Additional details regarding the derivation of the bound can be found in Appendix~\ref{sec:appendix_objective_function}.

Lastly, we address the potential practical advantage of defining the objective function in terms of NB/CNB regions as it pertains to reducing overhead required by QNS protocols. Noise characterization techniques, like QNS, are used to provide estimates of noise spectra by utilizing the quantum system as a dynamical probe. Such estimates can be critical to the design of noise-informed gates, as the noise suppression characteristics of the control are directly related to the spectral overlap between the FFs and the noise PSDs. Thus, in general, one requires reasonably sufficient characterization of the \emph{complete} noise PSD in order to design gates to minimize the overlap described in Eq.~(\ref{eq:overlap_general}). However, we find this condition to be too stringent and argue that it is sufficient to only require knowledge of key features, such as noise cutoff frequencies in order to define CNBs and estimates of the fractional power. Unconcerned with knowledge of the complete PSD, but rather just the ``flavor" of the noise, this approach potentially reduces the overhead required to provide sufficient estimates of noise PSDs via QNS.

\subsection{Gradients}
\label{sec:F-GRAFS-gradients}
In this section, we derive analytical gradient expressions for the objective function in Eq.~(\ref{eq:objective_function}) and the ideal gate fidelity $\mathcal{F}_G(T)$. Our derivation makes use of gradients originally introduced in Ref.~\cite{2018Lucarelli_GRAFS} for closed system DPSS-based optimized control. Note that similar analytical FF gradients have also been derived in Ref.~\cite{Le2022FilterOpt}. 

Under the parametrization of Eq.~(\ref{eq:ctrl-param}), where the DPSSs are selected to form the functional basis, the pure control evolution $U_C(t)$ is piece-wise constant. The control profiles resulting from the DPSS expansion inherit properties of the basis, namely, they are discrete sequences. Thus, for a given control sequence of $N$ timesteps, each of duration $\delta t$, the control amplitude will take constant values $\Omega_\nu(t)=\Omega_{\nu,n}$, where $t\in[t_n, t_{n+1})$ and $t_n = n\delta t$ for $n=0,1,\ldots, N-1$. Equivalently, projecting into the DPSS basis,
\begin{equation}
\Omega_{\nu,n} = \sum^{K-1}_{k=0} \alpha_{\nu, k}\, v^{(k)}_n,
\end{equation}
where we have dropped the explicit dependence on $N$ and $W$ for the DPSS. The piecewise-constant control assumption permits the control propagator $U_C(t)$ to be written as the product
\begin{equation}
U_C(t) = U_C(t, t_{n})\,\cdots \,U_C(t_2,t_1)U_C(t_1,0),
\label{eq:PWC_time_propagator}
\end{equation}
where $n=\lfloor t/\delta t\rfloor$.
Each constituent propagator $U_C(t_j,t_{j-1})$ implemented over the $j$-th time step is generated by the control Hamiltonian [Eq.~(\ref{eq:H_Ctrl})],
\begin{equation}
U_C(t_j,t_{j-1})=e^{-i\delta t\, \vOmega_j\cdot\vsigma/2}
\end{equation}
for $j=0,1,\ldots, N-1$ and $\vOmega_j=(\Omega_{x,j},\Omega_{y,j})$.

F-GRAFS optimizes the control waveform via optimization of the expansion coefficients $\{\alpha_{\nu,k}\}$. As a result, the coefficients are updated according to
\begin{equation}
\alpha_{\nu,k}^{(r+1)} = \alpha_{\nu,k}^{(r)} - \gamma \left.\frac{\partial}{\partial \alpha_{\nu,k}} \Gamma(T)\right|_{\alpha_{\nu,k}^{(r)}}
\end{equation}
at the $(r+1)$-th iteration of F-GRAFS. The initial values $\alpha^{(0)}_{\nu,k}$ can be chosen randomly or tailored to the noise characteristics, as we will discuss in detail in Sec.~\ref{sec:results}. The parameter $\gamma$ is the learning rate that is determined adaptively by the SLSQP algorithm.

The gradient of the objective function is proportional to the gradient of the FFs. The controls are finite duration and bounded by construction, and therefore the integral in Eq.~(\ref{eq:objective_function}) converges. Hence, the derivatives with respect to the expansion coefficients commute with the integral over frequencies and can be applied directly to the FFs as follows:
\begin{eqnarray}
\frac{\partial F_{\mu}(\omega)}{\partial \alpha_{\eta,k}} &=& \sum_{\nu} \frac{\partial R_{\mu \nu}(\omega)}{\partial \alpha_{\eta,k}} R_{\mu \nu}^*(\omega)
+ R_{\mu \nu}(\omega) \frac{\partial R_{\mu \nu}^*(\omega)}{\partial \alpha_{\eta,k}},\nonumber\\
\end{eqnarray}
for $\mu,\nu,\eta=x,y,z$. Note that we have dropped the explicit dependence on $T$ for brevity. By virtue of the Eq.~(\ref{eq:control_matrix_freq}) and subsequently, Eq.~(\ref{eq:control_matrix}), the derivatives propagate from the frequency-domain representation of the control matrices to its time-domain counterpart according to
\begin{eqnarray}
\frac{\partial R_{\mu\nu}(t)}{\partial \alpha_{\eta,k}} &=& \mathrm{Tr}\left(\frac{\partial U_C^\dagger(t)}{\partial \alpha_{\eta,k}}\sigma_\mu U_C(t)\sigma_\nu\right) + \nonumber\\
&& + \mathrm{Tr}\left(U_C^\dagger(t)\sigma_\mu \frac{\partial U_C(t)}{\partial \alpha_{\eta,k}}\sigma_\nu\right).
\end{eqnarray}
Employing the chain rule, and noting that $\frac{\partial\Omega_{\nu,n}}{\partial\alpha_{\eta,k}} = v_n^{(k)} \delta_{\mu\rho}$, we arrive at the derivatives 
\eq{
\label{eq:partials_Uc}
\frac{\partial U_C(t)}{\partial \alpha_{\eta,k}} = \sum_{n=0}^{N-1}\frac{\partial U_C(t)}{\partial \Omega_{\eta,n}} v_n^{(k)}
}
for the control propagator. We can again exploit the piecewise constant control assumption to determine the derivative of the control propagator with respect to the control amplitude. Letting 
\begin{equation}
    Q_{n:m}=U_C(t_{n+1},t_n)\,\cdots \,U_C(t_{m+1}, t_{m}) \nonumber
\end{equation}
denote the partial control propagator, the derivative can be expressed as
\begin{equation}
    \label{eq:Uc_gradient}
    \frac{\partial U_C(t)}{\partial \Omega_{\nu,n}} = Q_{N-1:n+1} \left(\frac{\partial}{\partial \Omega_{\nu,n}} U_C(t_{n+1}, t_n)\right) Q_{n-1:0}.
\end{equation}
Finally, each derivative of the controlled evolution during $t\in[t_n,t_{n-1})$ can be computed via exact diagonalization~\cite{1963aizu}, where the matrix elements of 
\eq{
\frac{d}{ds} e^{-i\delta t (A+sB)}
}
are given by
\eq{\label{eq:exp-partial}
\langle \lambda | &\frac{\partial e^{-i\delta t (A + sB)}}{\partial s} | \lambda^\prime \rangle = \\
&\begin{cases} \nonumber
-i\delta t \langle \lambda | B |   \lambda \rangle \, e^{-i\delta t \lambda} \quad &\mathrm{for} \, \, \lambda=\lambda^\prime \\ 
-i\delta t \langle \lambda | B |   \lambda^\prime \rangle 
\, \frac{e^{-i\delta t \lambda} - e^{-i\delta t \lambda^\prime} }{\lambda - \lambda^\prime}  &\mathrm{for} \, \, \lambda\neq\lambda^\prime
\end{cases}
}
where $\{|\lambda\rangle,\lambda\}$ is the eigensystem of the matrix $A$.

The final expression required is the gradient of the ideal gate fidelity $\mF_G(T)$. Using Eqs.~(\ref{eq:partials_Uc}) and (\ref{eq:Uc_gradient}), we find
\eq{
\frac{\partial \mF_G(T)}{\partial\alpha_{\nu,k}} &= \mathrm{Re} \left\{ \tr[U^\dagger_G U_C(T)]^* \tr\left[U^\dagger_G \frac{\partial U_C(T) }{\partial\alpha_{\nu,k}}\right] \right\}.
}
Together, the spectral leakage and ideal gate fidelity gradient expressions are used by F-GRAFS to generate controls for specified gate operations with tailored FFs. Below, we showcase the capabilities of F-GRAFS for a variety of noise and control scenarios.

\section{Noise-optimized gates}
\label{sec:results}
In this section, we demonstrate F-GRAFS's ability to discover noise-optimized controls in two different control and noise scenarios. First, we consider the case of single axis control along $\sx$ and dephasing noise only along $\sz$. In the second case, we study the more complex case of multi-axis control along $\sx$ and $\sy$, with dephasing noise along $\sigma_\mu$, $\mu=x,y,z$. In each subsection, we illustrate how to initialize F-GRAFS based on analytical expressions that can be tuned to the specifications of the CNB. Subsequent optimization is then used to produce optimized controls that simultaneously provide significant suppression of the FFs within the CNB and high-fidelity non-trivial single qubit operations. Lastly, we explore the relationship between the DPSS bandwidth parameter and the post-F-GRAFS residual spectral leakage. This analysis provides key insight into the interplay between the control bandwidth and the characterisics of the noise PSD.

\subsection{Single Axis Control and Dephasing}
We begin by studying a system driven via single axis control applied in the $x$ direction and subject to time-correlated dephasing noise along the $z$ axis. 
This scenario is compatible with the highly asymmetric case where the fractional power estimates are $\overline{p}_z=1$ and $\overline{p}_x=\overline{p}_y=0$.
At the Hamiltonian level, we impose $\Omega_y(t)=0$ and $\vec{\beta}(t)=(0,0,\beta_z(t))$ in Eqs.~(\ref{eq:H_Ctrl}) and (\ref{eq:H-noise}), respectively. Control along $\sx$ and noise along $\sz$ gives rise to two control matrix components $R_{yz}$ and $R_{zz}$ that ultimately contribute to $F_z(\omega)$, the only non-trivial FF for this case. The objective function [Eq.~(\ref{eq:objective_function})] therefore reduces to a single integral focused solely on the spectral leakage of $F_z(\omega)$ within the CNB $\overline{\mathcal{B}}_{z}$. Below, we consider two types of CNBs defining two distinct gate types: (1) \emph{high-pass gates}, where the noise is assumed to be low frequency and the CNB is described by a properly chosen high frequency cutoff dependent upon the characteristics of the dephasing noise PSD $S_z(\omega)$ and (2) \emph{band-pass gates}, where the noise PSD possesses both low and high frequency components. In the latter case, the CNB is determined by multiple cutoff frequencies to adequately capture the characteristics of $S_z(\omega)$.

\subsubsection{Analytically-Informed Initial Conditions}
\label{subsubsec:sa-ic}
The FF design problem constitutes a non-convex optimization problem that strongly depends on the initial conditions used for the gradient-based optimization procedure.
One approach is to randomly initialize the expansion coefficients $\alpha_{\nu,k}$; however, as we show in Appendix~\ref{sec:initial_conditions}, we find that this typically results in unstable solutions and unnecessarily large control amplitudes. We overcome this issue by utilizing primitive controls with straightforwardly and intuitively tunable FFs as initial conditions.

In particular, we employ constant drive (CD) control as an initial guess for the optimized control waveform. Known in noise characterization~\cite{willick2018cd} and quantum signal detection~\cite{titum2021optimal}, CD represents a simple control scheme in which the system is driven at a constant rate $\Omega(t) = \Omega_0$ for a time $T$. The amplitude $\Omega_0$ dictates the center frequency of the FF, while the total duration of the drive determines the spectral width; see Fig.~\ref{fig:SA_control_FF}(a) for example. Functionally simplistic, the CD FF $F(\omega,T)\propto T \sinc((\omega\pm\Omega_0))$, and in the limit $T\rightarrow\infty$, this FF converges to delta functions centered around $\pm \Omega_0$. The tunability and localization of the CD FF are key features that we exploit to initialize the F-GRAFS optimization.

Upon construction, the initial control waveform is projection into the DPSS basis and then optimized via F-GRAFS. First, the amplitude of the CD initial condition is determined by the CNB, while $T$ is dictated by the desired gate time. The subsequent (initial) control $\Omega^{(0)}_x(t)$ is then projected into the DPSS basis, where the expansion coefficients 
\eq{
\alpha_{x,k}^{(0)} = \sum^{N-1}_{n=0} \Omega^{(0)}_{x,n} v^{(k)}_n \delta t,
}
are determined via the DPSS orthogonality relation $\sum^{N-1}_{n=0} v^{(k)}_n v^{(k')}_n \delta t=\delta_{k,k'}$. F-GRAFS proceeds by optimizing these coefficients to reduce the spectral leakage of the FFs outside of the NB. CD affords some spectral concentration and reduced spectral leakage; however, as we will show in Sec.~\ref{subsubsec:optctrl-x}, F-GRAFS can further lessen residual leakage while abiding by the desired bandwidth constraints of the control.

\subsubsection{Optimized Control Waveforms and Filter Functions}
\label{subsubsec:optctrl-x}

\begin{figure}[t]
\centering
\includegraphics[width=\columnwidth]{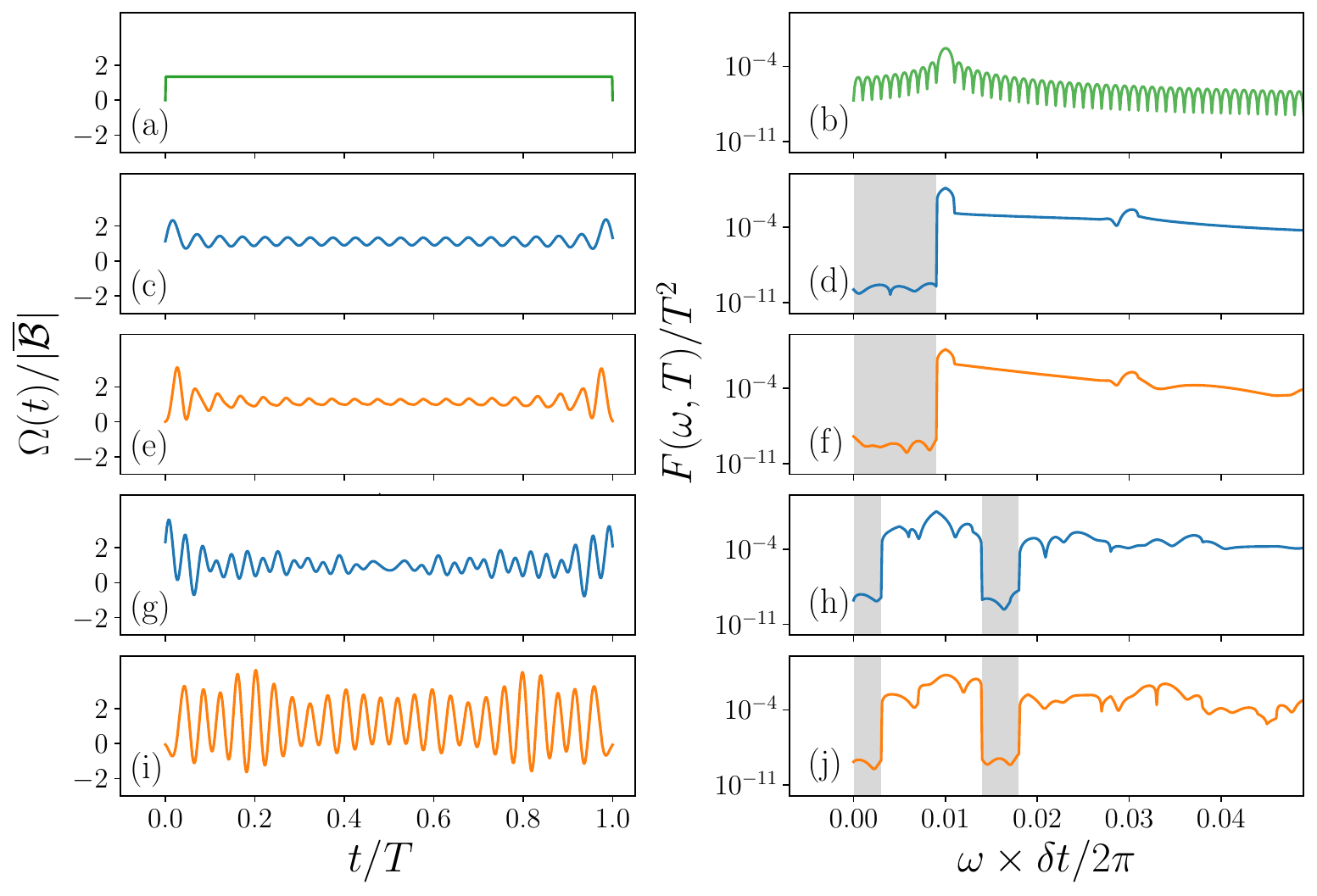}
\caption{
Control profiles generating $X_\pi$ gates (left) and their associated FFs (right) for single axis control and noise. The CD initial condition, with amplitude $\Omega(t)=\omega_H=0.01\times2\pi/\delta t$ and boundary conditions $\Omega(0)=\Omega(T)=0$ shown in (a) and (b). High-pass optimized control obtained from F-GRAFS and its associated FF shown in (c) and (d), respectively. The DPSS basis is specified by $K=2NW$, where $W=2\omega_H\times \delta t/2\pi$ and the CNB is $\overline{\mathcal{B}}=[0,\omega_H)$. Similar results are shown in (e) and (f) for an optimized waveform with endpoints near zero amplitude. The DPSS basis size is $K^\prime=2NW^\prime-4$, where $W^\prime=2W$. Band-pass optimized controls are shown in (g) and (i), with their associated FFs shown in (h) and (j). The passband is determined by $\overline{\mathcal{B}}=\overline{\mathcal{B}}_1\cup\overline{\mathcal{B}}_2$, where $\overline{\mathcal{B}}_1=[0,\omega_\ell)$ and $\overline{\mathcal{B}}_2=(\omega_\ell+\Delta\omega,\omega_H)$. The low-frequency cutoff is given by $\omega_\ell=0.004\times2\pi/\delta t$, the spectral width of the passband is $\Delta\omega=0.01\times2\pi/\delta t$, and the high-frequency cutoff of $\overline{\mathcal{B}}_2$ is $\omega_H=0.018\times 2\pi/\delta t$. Panels (g) and (h) use a DPSS of size $K$ and bandwidth $W=2 |\overline{\mathcal{B}}| \times \delta t/2\pi$, while panels (i) and (j) utilize $K^\prime$ DPSSs and $W^\prime=2W$ to impose near-zero control boundary conditions. All scenarios described above use a total number of $N=1000$ timesteps. In all cases, the optimized FFs attain several orders of magnitude improvement in cancellation within the CNB regions over CD.
}
\label{fig:SA_control_FF}
\end{figure}

Here, we illustrate the utility of F-GRAFS for constructing high-pass and band-pass gates using single axis control. Fig.~\ref{fig:SA_control_FF} serves as the focal point of this discussion, where we consider an $\pi$-rotation about the $x$-axis of the Bloch Sphere, $U_C(T)=X_\pi$, as a representative case. Control profiles are displayed in the left column, while the right column shows the corresponding FFs and CNBs (shaded regions). Each row indicates a different control and/or noise scenario. We begin with a discussion focused on how the initial conditions are informed by the CNB and then move to examining the optimized controls and FFs resulting from F-GRAFS. Optimizations are performed assuming a tolerance of $10^{-10}$ which typically requires $O(100)$ iterations of the SLSQP algorithm.

The amplitude of the CD initial condition is set by either the high frequency cutoff or the center frequency of the pass band. When the noise PSD predominately resides at low frequencies, the CNB is defined according to the high frequency cutoff $\omega_H$: $\overline{\mathcal{B}}=[0,\omega_H)$. Note that the size of a low frequency noise CNB equals its high-frequency cutoff, i.e. $|\overline{\mathcal{B}}| = \int_0^{\omega_H}d\omega=\omega_H$. Consequently, the amplitude of the CD is set to $\Omega^{(0)}_x(t)=\omega_H$ to center the FF at the edge of the NB; see Fig.~\ref{fig:SA_control_FF}(b). Note that any $\Omega^{(0)}_x(t)\geq \omega_H$ would suffice, however, centering a CD FF at higher frequency comes at the cost of higher amplitude controls. Furthermore, in practice, we find $\Omega^{(0)}_x(t)=\omega_H$ to be a sufficient initial condition for achieving an optimized control within approximately 100 iterations of F-GRAFS.

A similar approach is used when defining initial conditions for band-pass gates. Noise PSDs that have significant support at both low and high frequencies require CNBs to be defined as unions of disjointed regions in frequency space. For simplicity, we consider the case where there are two such regions: one at low frequency $\overline{\mathcal{B}}_1$ and another at intermediate frequencies $\overline{\mathcal{B}}_2$. The pass band existing between these CNBs can be defined according to a 
lower end $\omega_\ell$ and noiseless band of width $\Delta\omega$. In terms of these parameters, the CNB is given by $\overline{\mathcal{B}}=\overline{\mathcal{B}}_1\cup \overline{\mathcal{B}}_2$, where $\overline{\mathcal{B}}_1=[0,\omega_\ell)$ and $\overline{\mathcal{B}}_2=(\omega_\ell+\Delta\omega,\omega_H)$. Initial conditions defined from $\overline{\mathcal{B}}$ are dependent upon the lower cutoff frequency of the pass band, where $\Omega^{(0)}_x(t)=\omega_\ell$ can be used to initialize the optimizer. 

F-GRAFS offers improved suppression of the FF over CD within the CNB. In Fig.~\ref{fig:SA_control_FF}(c) and (d), the resulting F-GRAFS optimized control and FF, respectively, are shown for a high-pass gate. The normalized bandwidth of the DPSS basis is set to $W=2\omega_H\times \delta t/2\pi$. This choice is based on a bandwidth analysis of the FF suppression within the CNB; see Sec.~\ref{subsubsec:SA-optW} for further details. Comparing Fig.~\ref{fig:SA_control_FF}(a) and (c), we observe that while the CD offers some initial suppression of the FF within the CNB, F-GRAFS further reduces the FF contributions by approximately 4 orders of magnitude. Interestingly, the control profiles required to achieve this enhancement are qualitatively similar to the CD control. The distinctions lie in the high frequency oscillations centered about the CD-like control profile and the control boundaries $\Omega(0)$ and $\Omega(T)$ that differ from zero.

Through the DPSS basis, conditions on the endpoint to maximum control amplitude ratio can be imposed intrinsically, bypassing the need to append additional constraints to the objective function. From a practical perspective, control boundary conditions are desirable for ensuring the creation of viable optimized control profiles that abide by control hardware slew-rate limitations. Such conditions can be included in optimized control schemes via additional constraints~\cite{wilhelm2020introduction} or by analytically enforcing the constraints prior to optimization~\cite{Machnes2018Tunable}. The F-GRAFS approach essentially straddles the two approaches by imposing constraints on the DPSS basis elements prior to the optimization. More specifically, control boundary constraints can be indirectly enforced by imposing a minimum tolerance on the DPSS eigenvalues described in Eq.~(\ref{eq:slepian_eigen_eq}). The highest order DPSSs within $K\leq 2NW$ are the least spectrally concentrated and have nonzero amplitudes at the boundaries~\cite{2018Lucarelli_GRAFS}. By enforcing a spectral concentration constraint of $\lambda_k\geq \eta$, where $\eta$ is the desired tolerance, one can circumvent this issue and impose approximate boundary conditions on the DPSS basis. In order to maintain the same number of basis functions, the bandwidth $W$ must be artificially increased from $W$ to $W^\prime$. 

In practice, we find that boundary conditions can be sufficiently maintained by demanding $99\%$ concentration ($\eta=0.99$). Basis cardinality is preserved by doubling the normalized bandwidth $W^\prime=2W$, and taking the first $K^\prime=2\lfloor NW^\prime\rfloor-4$ DPSSs to form the new basis. An example of the control profiles and FFs resulting from the truncated basis optimization are shown in Fig.~\ref{fig:SA_control_FF}(e) and (f) for the high-pass filter case. FF suppression within the CNB is comparable to the results showing in panels (c) and (d). The control endpoint amplitudes are $\Omega(0)=0.01\times\omega_H$ and $\Omega(T)=0.05\times\omega_H$, compared to $\Omega(0)=1.13\times\omega_H$ and $\Omega(T)=1.32\times\omega_H$ obtained without the eigenvalue concentration constraint. Furthermore, we observe an improvement in spectral concentration. Controls obtained from the truncated basis with bandwidth $W^\prime$ possess greater than $99\%$ concentration within $(-W,W)$, while optimized control waveforms constructed with a DPSS basis of bandwidth $W$ typically achieve between $70-90\%$ concentration. Thus, F-GRAFS is able to achieve similar noise suppression capabilities, while approximately satisfying boundary conditions and attaining improved spectral concentration. Note that the endpoints can be further reduced by increasing spectral tolerance -- and increasing bandwidth, if one wishes to preserve basis cardinality.

F-GRAFS supports band-pass gate design for more complex noise scenarios. In Fig.~\ref{fig:SA_control_FF}(g), F-GRAFS optimized controls are displayed for the $X_\pi$ gate and the CNB $\overline{\mathcal{B}}=\overline{\mathcal{B}}_1\bigcup \overline{\mathcal{B}}_2$, as described earlier in this section. The spectral width of the pass band is $\Delta\omega=0.01\times 2\pi/\delta t$ and the high frequency cutoff of $\overline{\mathcal{B}}_2$ is $\omega_H=0.018\times 2\pi/\delta t$. The control bandwidth is set to $W=2|\overline{\mathcal{B}}|\times \delta t/2\pi$, where $|\overline{\mathcal{B}}| = \omega_H - \Delta\omega$. This choice is based on an analysis of the spectral leakage in the CNB as a function of bandwidth; see Sec.~\ref{subsubsec:SA-optW} for further elaboration. Noise suppression afforded by the optimized controls is depicted in Fig.~\ref{fig:SA_control_FF}(h), where the FF is shown to have spectral nulls within the CNB regions. In comparison to the CD initial condition, the F-GRAFS solutions achieve greater FF suppression; approximately four orders of magnitude improvement. Similar performance characteristics are observed when leveraging a truncated basis to satisfy control boundary conditions; see Fig.~\ref{fig:SA_control_FF}(i) and (j).

\subsubsection{Optimal Control Bandwidth for Single-Axis Noise}
\label{subsubsec:SA-optW}
The connection between control bandwidth and noise characteristics plays an important role in achieving noise-robust quantum gates. Control hardware is often subject to constraints on amplitude and bandwidth. In Sec.~\ref{subsubsec:sa-ic}, we alluded to the dependence of the FF on control amplitude when selecting initial conditions. Namely, the spectral cutoffs of the noise determined the control amplitude. Here, we examine the dependence of the FF on control bandwidth. In particular, we exploit the intrinsic tunability afforded by the DPSS to study the optimized spectral leakage in the CNB as a function of the DPSS bandwidth $W$.

\begin{figure}
\centering
\includegraphics[width=.45\textwidth]{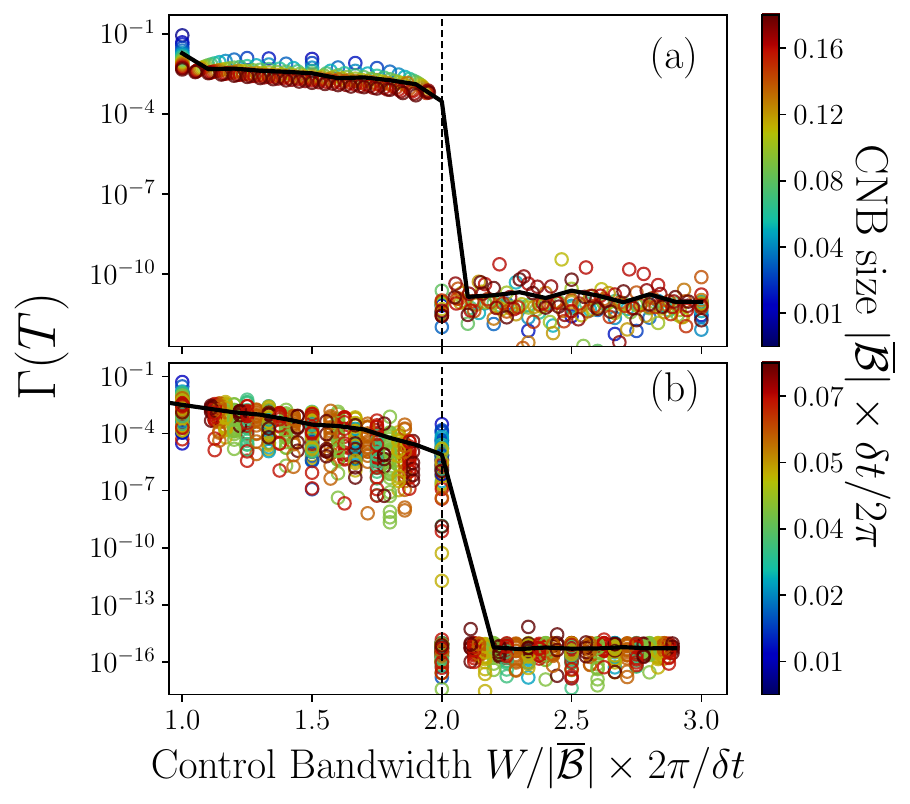}
\caption{
F-GRAFS optimized spectral leakage $\Gamma(T)$ as a function of the control bandwidth $W$ for single axis control and noise. Two noise scenarios are shown: (a) high-pass gates and (b) band-pass gates. Each colored data point corresponds to a different combination of $W$ and noise parameters, each averaged over five $X_\theta$ gates, where $\theta\in[0,\pi]$. In the high-pass case, the CNB size varies according to $|\overline{\mathcal{B}}|=\omega_H\in(0,0.16]\times2\pi/\delta t$. The band-pass case is specified by a CNB of size $|\overline{\mathcal{B}}|=\omega_H-\Delta\omega\in(0,0.08]\times2\pi/\delta t$. The number of different noise, control, and gate configurations totals (a) 2200 and (b) over 9000. In both cases, a transition occurs at the critical bandwidth $W_c=2|\overline{\mathcal{B}}|\times\delta t/2\pi$. For bandwidths $W<W_c$, the performance improves as $W$ increases, whereas for $W>W_c$ the performance of the optimization saturates to the gradient tolerance of (a) $10^{-12}$ and (b) $10^{-15}$. The solid black line represents the mean of the colored data points over bandwidth windows of size 0.1.
}
\label{fig:SA_bandwidth}
\end{figure}

In Fig.~\ref{fig:SA_bandwidth}(a) and (b), the spectral leakage $\Gamma(T)$ is shown as a function of bandwidth $W$ for the high-pass and band-pass cases, respectively. Data points represent different combinations of control bandwidth and noise parameters averaged over five different $X_\theta$ rotations, where $\theta\in[0,\pi]$. The high-pass comparison shown in panel (a) includes CNB sizes in the range $|\overline{\mathcal{B}}|=\omega_H\in(0,0.16]\times2\pi/\delta t$, resulting in a total of 2200 configurations. Panel (b) conveys similar results for the band-pass scenario, where $|\overline{\mathcal{B}}|=\omega_H-\Delta\omega\in(0,0.08]\times2\pi\delta t$ for the same $W$ range. A total number of 9000 configurations are included in the band-pass comparison. The significant increase in configurations results from the increased complexity of the band-pass scenario which in turn leads to a greater breadth in possible noise and control configurations one may consider. Data points are supplemented by the median of the data (black line) calculated using a bandwidth window of 0.1.

Numerical experiments indicate a uniform dependence of $\Gamma(T)$ on control bandwidth for both high-pass and band-pass gates. Both gate types exhibit a distinct phase transition in spectral leakage at the critical bandwidth $W_c=2|\overline{\mathcal{B}}|\times \delta t/2\pi$. Control bandwidths below the critical bandwidth result in significant spectral leakage. In contrast, we observe convergence in $\Gamma(T)$ for $W>W_c$, likely due to saturation of the designated optimization tolerance. Despite potential optimizer-dependent features at high control bandwidth, the emergence of the critical bandwidth appears to be generally only dependent on the size of the CNB. This seemingly universal behavior suggests that for a single qubit subject to single axis control and dephasing, the optimal control bandwidth is $W_c$.

Establishing an analytical justification for this relationship between $W_c$ and properties of the noise is challenging. 
This is primarily due to the non-linear relationship between the control and the FFs. That said, it is worth noting that the expression of $W_c$ is strongly reminiscent of the Nyquist-Shannon sampling theorem~\cite{landau1967nyq}. 
This theorem states that in order to effectively reconstruct a signal of a given bandwidth $B$, it suffices to sample at a frequency of $f_s=2B$. Without further constraints imposed on the signal, the theorem determines that this sampling frequency is both sufficient and necessary. 
It is within this context of classical signal reconstruction that we propose the following intuition. 
Rather than considering F-GRAFS as a method for optimally filtering noise, let us treat it as an approach for finding the controls required to ``reconstruct" an ideal FF that optimally filters noise within the CNB $\CNB$.
According to Fig.~\ref{fig:SA_bandwidth}, such a reconstruction requires a minimum control bandwidth of $W_c=2|\overline{\mathcal{B}}|\times \delta t/2\pi$ in units of $2\pi/\delta t$.
This value is, in turn, also the one needed to sample and reconstruct a signal of bandwidth $|\CNB|$.
This suggests that the problem of filtering a noise with a CNB of size $|\CNB|$ is equivalent (in the sense of resources needed) to sampling and performing signal reconstruction of a function of bandwidth $|\CNB|$.
 
The results obtained from F-GRAFS allow for an alternative interpretation, in terms of the Landau-Pollak theorem~\cite{landau1961prolate}. 
This theorem states that the dimensionality of a signal of bandwidth $W$ (in units of Hz) and total duration $T$ is $2WT$, or $2NW$ as expressed in the present units and quantities.
This is obtained by showing that the DPSSs can optimally approximate any time and bandlimited function of bandwidth $W$ with a DPSS basis of only $2NW$ elements.
Through F-GRAFS, we find that the optimal DPSS basis uses a bandwidth $W_c$ and $2NW_c$ elements. 
This basis is the one capable of approximating any time and bandlimited function of bandwidth $W_c$ as well, and F-GRAFS shows that the controls capable of noise filtering belong to this set of functions.
This seems hardly a coincidence: the amount of resources (degrees of freedom) needed to filter a noise with a CNB of size $|\CNB|$ are the same as the ones required to approximate a signal of bandwidth $2|\CNB|$, with effective dimensionality $4N|\CNB|$ as given by the Landau-Pollak theorem.
Let us reiterate that this signal of bandwidth $2|\CNB|$ is the one capable of filtering a noise with CNB size $|\CNB|$.

Based on these arguments, we claim that the FFs bridge the gap between the noise and the controls. Based on the sampling theorem, classically one could think that if the noise functions $\beta(t)$ reach a maximum bandwidth $\omega_H\geq|\CNB|$, the frequency $2\omega_H\geq W_c$ would have to act as an absolute minimum for the control functions bandwidth. The results from F-GRAFS show that the  non-linear FF transformation is able to capture the essential degrees of freedom of the noise needed for cancellation, compressing it into a space of dimension $2NW_c$, with $W_c=2|\CNB|\times \delta t/2\pi$.

\subsection{Multi-axis Control and Dephasing}
In situations where noise contributions are not limited to a single axis, F-GRAFS can be employed to simultaneously suppress noise along multiple axes. We illustrate this feature by considering two axis control, $\Omega_\nu(t)\neq0$, $\nu=x,y$, and noise along all three Pauli axes: $\vec{\beta}(t)=(\beta_x(t),\beta_y(t),\beta_z(t))$. We assume cylindrical symmetry and therefore require $\beta_x(t)=\beta_y(t)=\beta_{xy}(t)$. 
To simplify the analysis, in the present section we consider the symmetric case, where the fractional power estimates are equal $\overline{p}_x=\overline{p}_y=\overline{p}_z=1/3$, and vary the relative sizes of the CNBs.

In the absence of cross-correlations, six control matrix components give rise to three unique FFs $F_\mu(\omega)$ and PSDs $S_\mu(\omega)$, $\mu=x,y,z$. The objective function [Eq.~(\ref{eq:objective_function})] now contains three contributions over CNBs $\overline{\mathcal{B}}_\mu$. Below, we investigate the efficacy of F-GRAFS for arbitrary single qubit gates in both the high-pass and band-pass case.

\subsubsection{Analytically-Informed Initial Conditions}
\label{subsubsec:ma-ic}
The increased complexity in the control, FFs, and desired gate operations poses new challenges for the F-GRAFS optimization problem in the multi-axis setting. In particular, the ability of the algorithm to convergence on viable solutions is strongly dependent upon the initial conditions. In many cases, random initial conditions are not sufficient. Hence, we take an approach similar to the single-axis case and rely on analytically-informed initial conditions to improve algorithmic stability and convergence.

Initial conditions for multi-axis control optimization are constructed from the high-pass filtering scenario. More specifically, we consider the case where $\overline{\mathcal{B}}_x = \overline{\mathcal{B}}_y = [0,\omega_{xy})$ and $\overline{\mathcal{B}}_z = [0,\omega_{z})$. Initial conditions are derived based on a simplistic control ansatz: CD along one axis and a square wave along the second control axis. This approach is inspired by the single axis case and canonical dynamical decoupling sequences~\cite{ahmed2013robustness,lidar2013book} that leverage rapidly fluctuating control to mitigate noise. Despite its simplicity, this control ansatz proves to be amenable to more general scenarios beyond the conditions under which it is derived.

The analytical form of the initial conditions is most conveniently expressed in polar coordinates. The controls are thus represented as $\vOmega(t)=\Omega\,(\cos\phi(t),\sin\phi(t))$, where $\phi(t)=\phi_0 s_\lambda(t)$. The modulation function $s_\lambda(t) = \pm1$ defines a square wave with unit amplitude and control frequency $\lambda=2\pi/T_c$. $T_c$ is the control period, and it satisfies $T=MT_c$ for some positive integer $M$. An example of this control is given in Fig.~\ref{fig:MA_CDSW_control_FF}(a). 
The values of the control parameters required to filter low frequency noise up to $\omega_z=n_z\delta\omega$ and $\omega_{xy}=n_{xy}\delta\omega$ are found by setting $M,\theta$ as defined below and solving the following system of equations
\eq{
\begin{cases}
M &= n_z + n_{xy}, \\
\theta &= 2\pi n_z/M, \\
\frac{\sin^2(\xi)}{1-\xi\cot(\xi)} &= \cos^2(\theta/4), \\
\xi &= \frac{\Omega T}{2M},\\
\tan^2\phi_0 &= -\xi \cot(\xi),
\end{cases}
\label{eq:ma-init-conds}
}
Solving the third equation numerically using standard optimization libraries, one finds a family of potential parameter choices. Motivated by efficient use of resources, we select the solution that minimizes $\xi$ and as a consequence the control amplitude. Note that much like the single axis case, this control ansatz affords intuition in parameter selection. Namely, the control amplitude is proportional to the sum of the noise cutoff frequencies $M$, up to a factor determined by $\xi$. Furthermore, the single-axis initial control ansatz can be recovered by setting $\omega_{xy}=0$, which in turn yields $\phi_0=0$ and single axis CD control. Further details on the derivation of the system of equations given in Eq.~(\ref{eq:ma-init-conds}) can be found in Appendix~\ref{sec:MA_IC}.

An example of initial conditions for the controls and their associated FFs are shown in panels (a) and (b) of Fig.~\ref{fig:MA_CDSW_control_FF}, respectively. Note that $\Omega_x(t)$ is chosen to drive the single qubit system according to a CD, while $\Omega_y(t)$ utilizes the fluctuating square wave. Results are shown for $\omega_z=\omega_{xy}=0.02\times 2\pi/\delta t$. For this specific case of noise parameters, $n_{xy}=\omega_{xy}/\delta\omega=n_{z}=\omega_{z}/\delta\omega=8$. This yields controls parameter values of $M=16$ and $\theta=\pi$. Additionally, we obtain the smallest value of $\xi\approx1.937$ after solving the third equation in Eqs.~(\ref{eq:ma-init-conds}) and $\phi_0\approx0.711$ after inverting the last one. 

\subsubsection{Optimized Control Waveforms and Filter Functions}
\begin{figure}[t]
\centering
\includegraphics[width=\columnwidth]{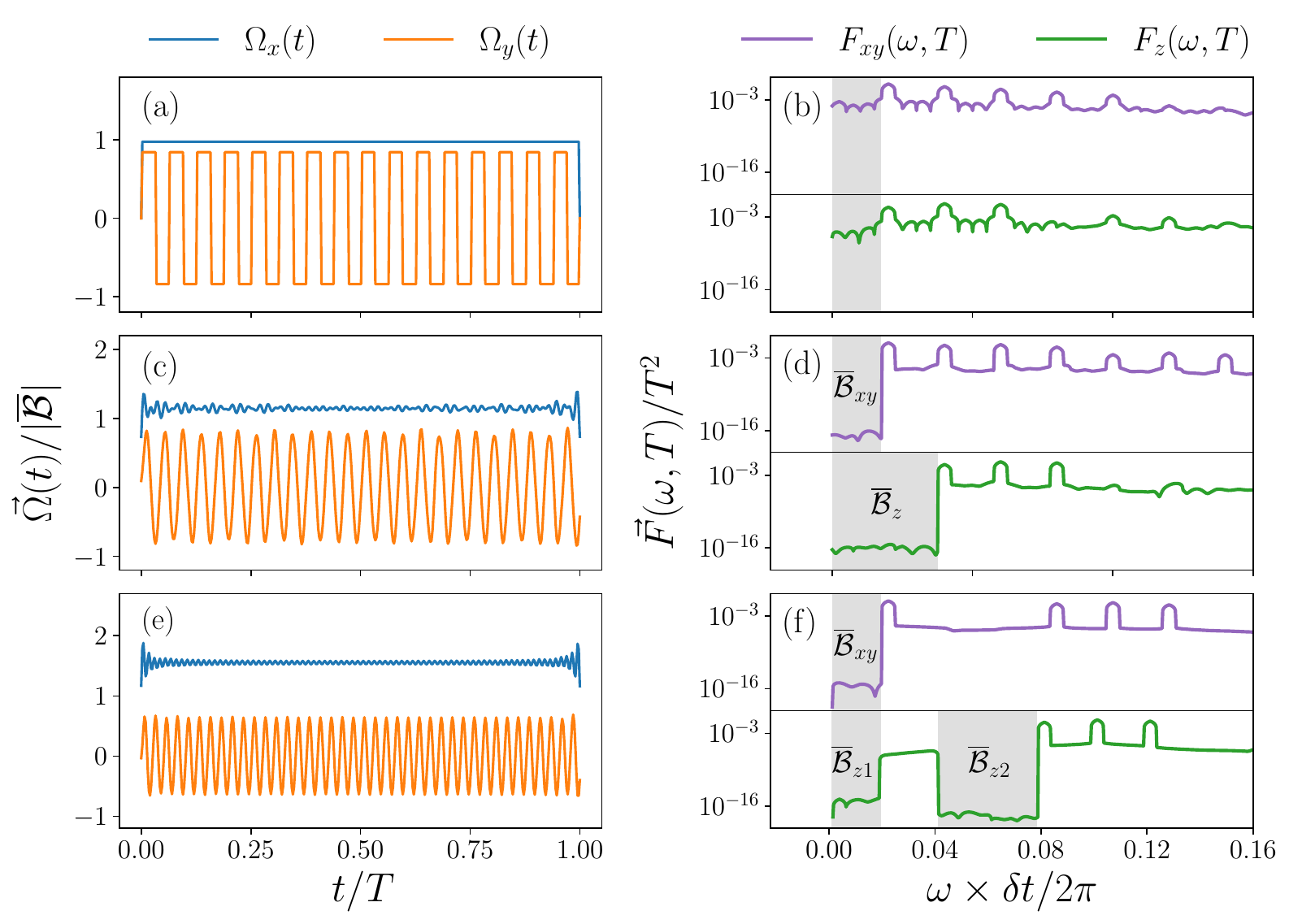}
\caption{
F-GRAFS optimized controls and FFs for multi-axis control and noise. The left column conveys control profiles, where each panel shows control waveforms for $\Omega_x(t)$ (blue line) and $\Omega_y(t)$ (orange line). The right column depicts the corresponding FFs, with the top of half of the panel displaying $F_{xy}(\omega,T)=F_x(\omega,T)+F_y(\omega,T)$ and the bottom half containing $F_z(\omega,T)$. The analytical initial conditions are shown in (a) and (b) for $\lambda=\omega_z+\omega_{xy}=0.04\times2\pi/\delta t$. F-GRAFS optimized (c) controls and (d) FFs for the high-pass case are shown for $\overline{\mathcal{B}}_{xy}=[0,\omega_{xy})$ and $\overline{\mathcal{B}}_{z}=[0,2\omega_{xy})$, where $\omega_{xy}=0.02\times2\pi/\delta t$. Similarly, F-GRAFS optimized (e) controls and (f) FFs for the band-pass case are shown for $\overline{\mathcal{B}}_{xy}=[0,\omega_{xy})$ and $\overline{\mathcal{B}}_z=\overline{\mathcal{B}}_{z1}\cup\overline{\mathcal{B}}_{z2}$. The CNBs for the band-pass regions are $\overline{\mathcal{B}}_{z1}=[0,\omega_{xy})$ and $\overline{\mathcal{B}}_{z2}=(\omega_{xy}+\Delta\omega,\omega_{z})$. $\omega_{xy}$ is the same as in the high-pass case, while $\Delta\omega=\omega_{xy}$ and $\omega_z=4\omega_{xy}$. In all cases, $N=1000$ timesteps are used and the control bandwidth is $W=2|\overline{\mathcal{B}}|\times \delta t/2\pi$. As in the single axis noise case, the resulting optimized FFs are suppressed by several orders of magnitude with the CNBs.
}
\label{fig:MA_CDSW_control_FF}
\end{figure}

F-GRAFS optimized control waveforms and FFs for the multi-axis case are displayed in Fig.~\ref{fig:MA_CDSW_control_FF}. The left column contains the controls, while the right column shows the FFs. All cases perform the optimization of the same arbitrary single qubit gate using $N=1000$ timesteps. We illustrate the utility of F-GRAFS in two scenarios: the fully high-pass case and a hybrid band-pass/high-pass case. The former showcases F-GRAFS' ability to uniquely tailor FFs based on distinct CNBs and therefore noise properties. The latter further conveys this message with a more complex noise scenario. 

Here, we show that F-GRAFS can provide significant suppression of multi-axis, non-uniform low-frequency noise. Prior to optimization, the initial conditions are set according to the procedure described in the previous section. Thereafter, the controls are projected into the DPSS basis using a control bandwidth $W=2|\overline{\mathcal{B}}|\times \delta t/2\pi$. We elaborate on this choice in bandwidth in Sec.~\ref{subsubsec:ma-optw}. In Fig.~\ref{fig:MA_CDSW_control_FF}, panels (c) and (d), we show F-GRAFS optimized controls and FFs for a particular non-uniform high-pass gate scenario. Panel (d) includes $F_{xy}(\omega,T)=F_x(\omega,T) + F_y(\omega,T)$ in the top half of the plot, while the lower half displays $F_z(\omega,T)$. The example scenario is described by a CNB $\overline{\mathcal{B}}_{xy}=[0,\omega_{xy})$, where $\omega_{xy}=0.02\times 2\pi/\delta t$, and $\overline{\mathcal{B}}_z=[0,\omega_{z})$, with $\omega_z=2\omega_{xy}$. Optimized control profiles maintain much of the qualitative features of the   initial conditions. Yet, through small, optimized fluctuations in the controls, F-GRAFS controls yield many orders of magnitude improvement in FF suppression within the CNBs.

Lastly, we explore the hybrid case where optimized controls generate both high-pass and band-pass FFs. We consider the case where $\overline{\mathcal{B}}_{xy}=[0,\omega_{xy})$ and $\overline{\mathcal{B}}_z=\overline{\mathcal{B}}_{z1}\cup \overline{\mathcal{B}}_{z2}$; thus, requiring high-pass filtering along $\sx$ and $\sy$ and band-pass filtering along $\sz$. The CNB $\overline{\mathcal{B}}_{xy}$ is bounded by  $\omega_{xy}=0.02\times2\pi/\delta t$, with 
the low-frequency CNB for the $\sz$ channel also being determined by $\omega_{xy}$: $\overline{\mathcal{B}}_{z1}=[0,\omega_{xy})$. The passband is chosen to reside between $\overline{\mathcal{B}}_{z1}$ and $\overline{\mathcal{B}}_{z2}$, where $\overline{\mathcal{B}}_{z2}=(\omega_{xy}+\Delta\omega,\omega_{z})$. The high-frequency CNB is characterized by the width $\Delta\omega=\omega_{xy}$ and high frequency cutoff $\omega_{z}=4\omega_{xy}$. Upon optimization, we find that F-GRAFS offers substantial FF suppression within the CNBs; again, approximately ten orders of magnitude.

In examining the optimized control waveforms in Fig.~\ref{fig:MA_CDSW_control_FF}(c) and (e), a notable observation is the apparent resemblance between $\Omega_y(t)$ and a sinusoidal function. As we show discuss in Appendix~\ref{sec:MA_IC}, one can consider an alternative initial control ansatz, where the square wave is replaced with a sine function: $\Omega_\nu(t)=A_\nu + B_\nu\sin(\lambda t)$. Although lacking an analytical proof of its effectiveness, it can be shown numerically to perform equally as well as the initial conditions presented in Sec.~\ref{subsubsec:ma-ic}. 

\subsubsection{Optimal Control Bandwidth for Multi-axis Noise and Control}
\label{subsubsec:ma-optw}

\begin{figure}[t]
\centering
\includegraphics[width=.45\textwidth]{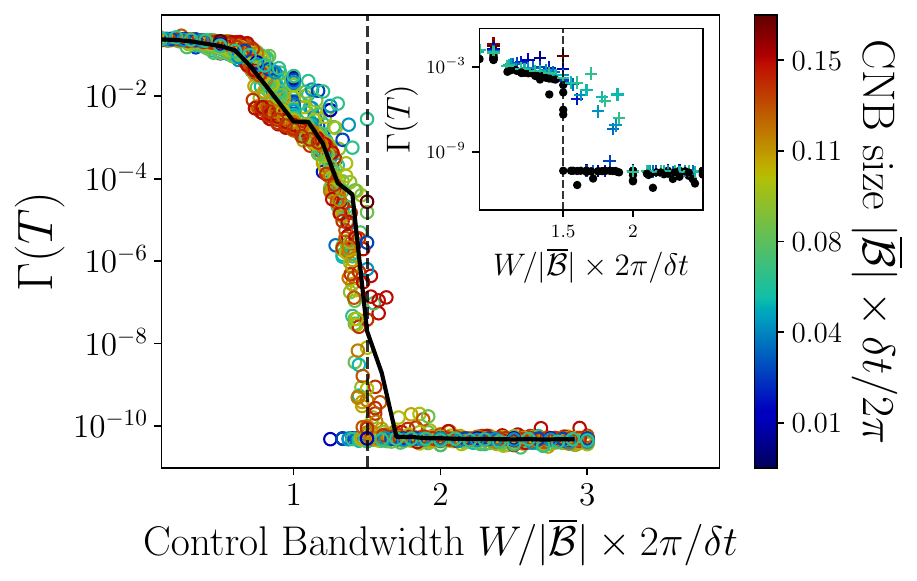}
\caption{F-GRAFS optimized spectral leakage $\Gamma(T)$ as a function of the control bandwidth $W$ for multi-axis control and noise. Plot depicts results from a high-pass scenario parametrized by spectral cutoffs $\omega_{xy}$ and $\omega_z$. As a result, the size of the CNB is given by $|\overline{\mathcal{B}}|=\omega_z+2\omega_{xy}$. Each colored data point corresponds to a distinct combination of $W$ and noise spectral cutoffs, resulting in CNB varying according to $|\overline{\mathcal{B}}|\in(0,0.16]\times2\pi/\delta t$. Data points are averaged over the seven Clifford+$T$ gates for a total of over 13000 different noise and control configurations. Note that a transition occurs at the critical bandwidth $W_c=1.5\times|\overline{\mathcal{B}}|\times\delta t/2\pi$. For bandwidths $W<W_c$, the performance improves as $W$ increases, whereas for $W>W_c$ the performance of the optimization saturates to the gradient tolerance below $10^{-10}$. The solid black line represents the mean of the colored data points over bandwidth windows of size 0.1. Inset: Single axis noise configurations with $\omega_{xy}=0$, solved with multi-axis control F-GRAFS. Colored crosses (black dots) represent the mean (minimum) values of optimized $\Gamma(T)$ over the Clifford+$T$ set.
}
\label{fig:MA_bandwidth}
\end{figure}

Here, we investigate the relationship between the properties of multi-axis noise and the control bandwidth. In Fig.~\ref{fig:MA_bandwidth}, the spectral leakage $\Gamma(T)$ is shown as a function of normalized control bandwidth. We consider 13000 high-pass gate scenarios using non-uniform CNBs and the Clifford+$T$ gate set $\mathcal{S}=\{I,X,Y,Z,H,S,T \}$ as the desired operations. As in the single-axis case, a strong relationship between control bandwidth and the high-frequency cutoffs is observed. More specifically, $\Gamma(T)$ decreases with increasing bandwidth, where the most rapid decline occurs near the critical frequency $W_c=1.5\times|\overline{\mathcal{B}}|\times \delta t/2\pi$ where $|\overline{\mathcal{B}}|=\omega_z+2\omega_{xy}$. The abrupt transition thereafter manifests due to the saturation of the optimizer to the specified gradient tolerance. Note that this critical proportionality factor of 1.5 between $W$ and $|\overline{\mathcal{B}}|\times \delta t/2\pi$, is lower than the critical value of 2 found in Sec.~\ref{subsubsec:SA-optW} for the case of single axis noise and control. 

We conjecture that the reduction in the optimal bandwidth condition is due to the additional degree of freedom in the control. As such, we investigate the dependence of $\Gamma(T)$ on $W$ by reducing the noise degrees of freedom to single-axis dephasing along $\sz$ and maintaining multi-axis control. The initial conditions for multi-axis control are determined by the properties of the noise and return to the single-axis CD case along $\sx$ when $n_{xy}=0$. Despite the single-axis initial condition, the optimizer has freedom to activate the control along $\sy$. Optimized $y$ control are in general non-zero, but typically remain smaller in amplitude than the optimized $x$ control.
The inset in Fig.~\ref{fig:MA_bandwidth} shows the results of the F-GRAFS optimization for the single-axis noise and multi-axis control setting using the $\mathcal{S}$ gate set. Black dots represent the minimum spectral leakage over the gate set, while the crosses represent the mean values.  Interestingly, the mean values revert to the single-axis noise and control critical bandwidth with a proportionality factor of 2, with minimum values being consistent with a proportionality factor of 1.5. Examining the control profiles, we find that the optimized controls more closely resemble single-axis optimized control along $\sx$, with a small fluctuating component along $\sy$.

Control power remains relatively constant despite the additional control degree of freedom. A reduction in control bandwidth could imply an increase in an alternative control resource, such as control power. In order to eliminate this possibility, we investigate the dependence of the optimized control power on control bandwidth. We find that no distinguishing features appear for $W>W_c$. Furthermore, the power of the optimized controls remains close to those of the initial conditions. This suggests that the improvement in performance is not provided by an increase in control amplitude, but rather due to the additional availability of control along $\sy$; see Appendix~\ref{subsec:MC_SN_power} for further details. 

Although the increased degree of control appears to play a role in determining the critical control bandwidth, there are other factors within the optimization problem that can also alter $W_c$. The F-GRAFS optimization problem is parameterized by the noiseless gate fidelity tolerance $\epsilon_G$. While $W_c$ does not appear to vary with $\epsilon_G$ in the single-axis control case, we observe dependence on this parameter in the multi-axis case. In particular, increasing $\epsilon_G$ in the SLSQP optimization facilitates a reduction in $W_c$ from 2 to 1.5 times the size of the CNB for the multi-axis control and single-axis noise scenario. The magnitude of the spectral leakage does not suffer from the lower fidelity tolerance, and degradation in ideal gate fidelity appears to be rather negligible. Thus, we suspect that the critical bandwidth coinciding with the single-axis control is due merely to the optimizer rather than being an intrinsic property of the control problem.

An additional interesting feature of the multi-axis control setting is that $W_c$ is gate dependent. By adjusting the fidelity tolerance, one can achieve a near-1.5 proportionality for single-axis noise for a subset of gates in $\mathcal{S}$. Among the gates that typically require bandwidth closer to twice the CNB is the identity. An adjustment of the initial conditions can yield reductions in critical bandwidth for the identity gate as well as other gates; however, the subset of gates that convey bandwidth improvements is predominately initial-condition-dependent. This behavior indicates that while the initial conditions shown here possess intuitive features, they are not universally favorable for all gates.

\subsection{On Optimal Bandwidth and Reachability}
The relationship between the characteristic timescales of the control and noise has been key to understanding the effectiveness of a control strategy in quantum control. For example, in pulse-based dynamical error suppression, the typical statement is: the interaction between the system and its environment can be effectively averaged out by utilizing pulses with inter-pulse delays much shorter than the characteristic timescale (or equivalently, the inverse of the high frequency cutoff) of the noise ~\cite{Viola1999DD,uys2009:lodd, khodjasteh2011:badd}. Albeit qualitatively instructive, this guiding principle is quite nebulous in that it is specific to ideal, instantaneous, pulse-based schemes and does not encompass more generic, smooth control. Moreover, it does not speak to optimality when striving to minimize control resources (such as bandwidth and power) while maximizing the effectiveness of the control. 

The numerical studies in Secs.~\ref{subsubsec:SA-optW} and \ref{subsubsec:ma-optw} address these issues and provide quantitative insight into the interplay between control bandwidth -- in the multi-axis control case, control power as well -- and the spectral properties of the noise. Empirical bounds enable the identification of optimal control bandwidth conditions for a variety of single and multi-axis control and noise scenarios, including those where the noise has significant spectral support at low and high frequencies. As a result, we supplement general criteria for pulse-based error suppression with explicit conditions that apply to a wide range of smooth control strategies and complex noise environments.

Furthermore, our numerical analysis of optimal bandwidth speaks to notions of reachability when subject to limited control resources. In control theory, reachability refers to the ability to drive a system from a given initial state to a set of final states, i.e., a reachable set. Equivalent notions of reachability have been developed in the quantum domain, where the reachable set can be described by a set of achievable unitaries~\cite{Wu2015reach, arenz2017reach}. Gate fidelity measures such as Eq.~(\ref{eq:op-fidelity}) are commonly used to quantify distance between the target and controlled unitaries and determine reachable sets that can be achieved within a specified tolerance~\cite{wu2011ctrl, arenz2017reach}.  The connection between Eq.~(\ref{eq:op-fidelity}) and the spectral leakage via the FFF suggests that $\Gamma(T)$ can act as a proxy for investigating reachability. It is within this context that we associate the minimum spectral leakage with attaining the reachable set. Thus, we find empirical evidence for saturation in reachability for control bandwidths beyond $W_c$ in both the single-axis and multi-axis control settings. The reachable set for single-axis control corresponds to arbitrary $X$ rotations, while $\mathcal{S}$ serves as the reachable set for multi-axis control. Note that in the latter case, by the Solovay-Kitaev theorem~\cite{nielsen00}, the reachable set provides access to the full SU(2) group and therefore, speaks to notions of controllability with limited control resources as well. Lastly, we note that while this approach does not supply rigorous analytical insight, it can be quite informative for identifying regimes where one expects to achieve a reachable set of logic operations with high fidelity when subject to control bandwidth constraints and a variety of control and noise scenarios.

\section{F-GRAFS Efficacy in Simulations}
\label{sec:simulation}
In Sec.~\ref{subsec:opt-prob}, the F-GRAFS optimization problem is defined through the spectral leakage as opposed to a distance metric. However, we argue that solving the F-GRAFS problem can be viewed as optimizing the upper bound on the phase invariant distance $D(U_G,U(T))$ in Eq.~(\ref{eq:dist-bound}). We substantiate this claim in this section by comparing the upper bound calculated through the F-GRAFS objective function to full dynamics simulations of a noisy single qubit driven by optimized control.

The efficacy of F-GRAFS is examined for a single qubit subject to multi-axis additive dephasing. Each noise component $\beta_\mu(t)$, $\mu=x,y,z$, is defined as a Ornstein-Uhlenbeck (OU) process. The PSD of the OU process is given by
\begin{equation}
    S_{OU}(\omega)=\frac{2\sigma^2\gamma}{\gamma^2+\omega^2},
\end{equation}
where $\sigma$ denotes the standard deviation and the parameter $\gamma$ is effectively related to the correlation time of the noise $\tau\sim1/\gamma$. For simplicity, we assume uniform noise along all three Pauli channels, i.e., $\beta_\mu(t)$ is generated by a process with an equivalent standard deviation and correlation time for all $i$. Note that cross-correlations are not permitted by construction, a condition enforced through this study. Each noise process is simulated by $\beta_{n+1} = (1-\gamma \delta t) \beta_n +\sigma\sqrt{2\gamma}w_n$, where $w_n$ and $\beta_0$ are drawn from normal distributions, with variance $\sqrt{\delta t}$ and $\sigma$, respectively. The parameter $\delta t$ again denotes the resolution of the control.

The control bandwidth and CNB are determined by the spectral features of the noise. The OU process defines a Lorentzian spectrum that is primarily concentrated at low-frequency. As such, the objective of F-GRAFS is to engineer high-pass gates with minimal spectral support in the CNB $\overline{\mathcal{B}}_\mu=[0,\omega_H)$, $\mu=x,y,z$. A relationship between the high-frequency cutoff $\omega_H$ and $\gamma$ can be determined analytically through explicit integration of the PSD. Denoting the total noise power as $P(\sigma)=\int_{0}^{\infty} S_{OU}(\omega)d\omega=\pi\sigma^2$, it can be shown that the fractional power in the CNB $\int_{\overline{\mathcal{B}}_\mu} S_{OU}(\omega)d\omega/P = 1-\epsilon$ can be used to derive $\omega_H=\gamma\tan[(1-\epsilon)\pi/2]$. We demand that $99\%$ of the noise power be concentrated within the CNB and therefore choose $\epsilon=0.01$ for the optimization. Note that the specifications of $\omega_H$ also dictate the optimal DPSS bandwidth $W=2|\overline{\mB}|=6\omega_H$ used in this example. 

\begin{figure}[h!]
\centering
\includegraphics[width=.45\textwidth]{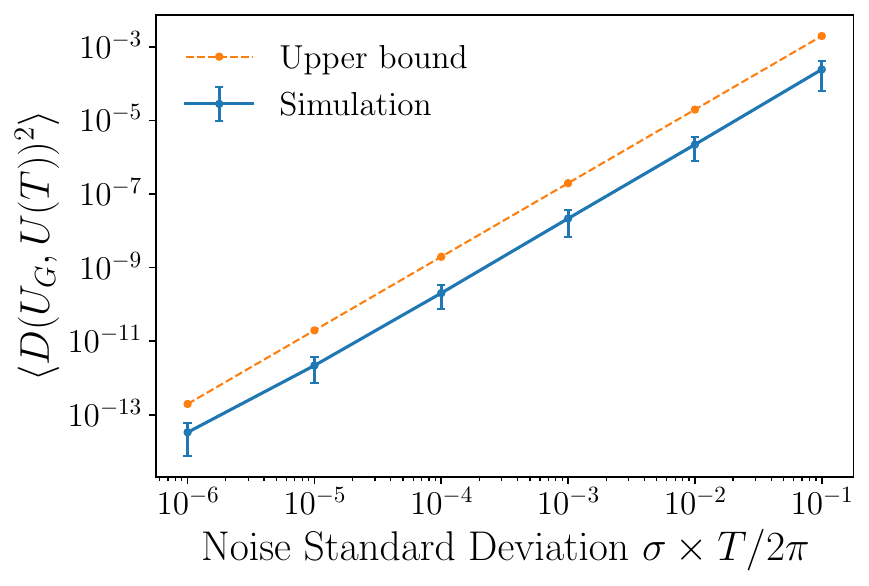}
\caption{Comparison between the upper bound informed by F-GRAFS and full dynamics simulations for multi-axis control and noise. Solid lines indicate simulations of noise filtering using Lorentzian noise created by an OU process and optimized control functions generated by F-GRAFS.
Results are shown for uniform noise along all three Pauli channels varied from low and high-frequency regimes according to $\gamma=[0.1,1.]\times2\pi/T$, with $\sigma\in[10^{-6},10^{-1}]\times2\pi/T$ controlling the noise power. The CNBs $\overline{\mathcal{B}}_\mu=[0,\omega_H)$ are determined by $\omega_H=\gamma\tan((1-\epsilon)\pi/2)$, where $\epsilon=0.01$ is chosen to allow only $1\%$ spectral leakage of the noise outside of the CNB. Dashed lines denote the upper bound [Eq.~(\ref{eq:dist-bound})] using F-GRAFS optimized controls. Control waveforms are chosen to have $N=128$ timesteps. F-GRAFS optimization is performed with a gate fidelity tolerance $\epsilon_G = 10^{-15}$ and gradient tolerance $10^{-20}$.
}
\label{fig:MA_simulation}
\end{figure}

Confirmation of the upper bound for F-GRAFS optimized controls is displayed in Fig.~\ref{fig:MA_simulation}. Full dynamics simulations are averaged over 1000 noise realizations and the single qubit Clifford+$T$ gate set. The solid lines correspond to the averages of the phase invariant distance, computing $U(T)$ [Eq.~(\ref{eq:time_propagator})] from simulations. Dashed lines denote the upper bound from Eq.~(\ref{eq:dist-bound}) computed using the F-GRAFS minimized spectral leakage. The discrepancy between the curves, and therefore the tightness of the bound, is dictated by omitted contributions from both $\mathcal{K}(T)$ and the $1\%$ leakage outside of the CNB; see Appendix~\ref{sec:appendix_objective_function} for further insight. 

The upper bound is maintained over a wide range of noise powers, justifying the F-GRAFS approach. As a surrogate objective function for the phase invariant distance, the spectral leakage (subject to an ideal gate constraint) proves to be sufficient for designing temporally correlated noise-robust gates for single qubit systems. Furthermore, the upper bound supports the use of NBs/CNBs rather than the complete noise spectrum. This observation provides an alternative perspective and potential focus for quantum noise spectroscopy protocols that may substantially reduce the typical cost of estimating the full noise PSD.

\section{Conclusion}
\label{sec:conclusion}
In summary, we have introduced a method for optimizing control in the presence of temporally correlated noise based on the FFF. Known as F-GRAFS, this approach seeks to simultaneously tailor FFs to minimize spectral overlap with a noise PSD and achieve non-trivial single qubit operations. Motivated by the need for improved F-GRAFS algorithmic convergence, we develop analytical control ansatze that are intuitively tunable based on the spectral cutoffs of the noise PSD. Their structure, albeit simplistic, is highly versatile and applicable to a variety of multi-axis noise scenarios. Furthermore, these analytical control schemes prove to be key to achieving fast F-GRAFS algorithmic convergence. 

F-GRAFS accommodates practical limitations in control hardware through the use of the DPSS basis. Characterized by an intrinsic bandwidth parameter, the DPSS provide a natural approach to constructing optimized controls that inherently abide by control hardware restrictions. We show that F-GRAFS can produce optimized control waveforms that significantly reduce spectral support of the FFs in designated frequency bands, while maintaining spectrally concentrated control. 

Together, F-GRAFS and the DPSS basis provide key insights into the connection between control parameters, noise characteristics, and optimal FF design. We leverage the intrinsic tunability of the DPSS to examine the noise suppression capabilities of optimized control protocols as a function of the DPSS bandwidth. We show that in both single axis and multi-axis noise scenarios there exists an identifiable optimal bandwidth proportional to twice and one-half the size of the region over which the FF is to be suppressed, respectively. 

Follow-on work would focus on providing an analytical understanding of the optimal control bandwidth condition and extending F-GRAFS to the multi-qubit regime. The non-linear relationship between the control and FF poses challenges for analytically deriving the optimal bandwidth condition. However, an analytical proof may shed light on key features of optimal control in the presence of temporally correlated noise processes. Extensions of F-GRAFS beyond the single qubit case could aid in expanding and generalizing the relationship between optimal control and the spectral properties of the noise. 

\section{Acknowledgements}
Y.O., K.S., B.D.C, and G.Q. acknowledge support from the U.S. Department of Energy, Office of Science, Office of Advanced Scientific Computing Research, Accelerated Research in Quantum Computing under Award Number DE-SC0020316. K.S. and G.Q. acknowledge support from ARO MURI grant W911NF-18-1-0218. D.L., K.S., B.D.C., and G.Q acknowledge support from the Intelligence Advanced Research Projects Activity via
Department of Interior National Business Center Contract No.~2012-12050800010. The U.S. Government is authorized to reproduce and distribute reprints for Governmental purposes notwithstanding any copyright annotation thereon. The views and conclusions contained herein are those of the authors and should not be interpreted as necessarily representing the
official policies or endorsements, either expressed or implied, of IARPA, DoI/NBC, or the U.S. Government.

\appendix

\section{Objective function}
\label{sec:appendix_objective_function}
In this section, we derive the F-GRAFS objective function.
As described in the main text, to quantify the  performance of the optimized gates, we use the phase-invariant distance Eq.~(\ref{eq:op-dist}). 
Taking the square and using the triangle inequality, the following upper bound on the average squared-distance can be established:
\begin{eqnarray}
\braket{D(U_G,U(T))^2} &\leq& \braket{\left( D(U_G,U_C(T)) + D(\id ,\tU(T)) \right)^2} \nonumber\\
&\leq& 2-\mathcal{F}_G(T) - \mathcal{F}_N(T) + \mathcal{K}(T),
\label{eq:dist-bound}
\end{eqnarray}
where $\langle\cdot\rangle$ denotes an average over noise realizations. $\mathcal{K}(T)$ is the cross term resulting from expanding the square. 

We now focus our attention on the average noise fidelity $\mF_N(T)$ and relate it to the spectral leakage $\mF_\Gamma(T)$. Through Eq.~(\ref{eq:fidelity_overlap}), $\mF_N(T)$ can be related to the overlap $\chi(T)$. As seen in Eq.~(\ref{eq:overlap_general}), $\chi(T)$ is defined as the overlap integral between the PSDs and the FFs, summed over all axes $\mu=x,y,z$
\eq{
\label{eq:discrete_overlap}
2\pi\chi(T) &= \sum_{\mu=x,y,z} \int_0^\infty S_\mu(\omega) F_\mu(\omega) d\omega \nonumber\\
&= \sum_{\mu=x,y,z} \left( \int_{\overline{\mathcal{B}}_\mu}  + \int_{\mathcal{B}_\mu} \right) S_\mu(\omega) F_\mu(\omega) d\omega \nonumber \\
&= \sum_{\mu=x,y,z} \left( \sum_{n\in\overline{\mathcal{N}}_\mu} + \sum_{n\in\mathcal{N}_\mu} \right) S_{\mu,n} F_{\mu,n} \delta\omega.
}
In the last line, we imposed the restriction that since the total pulse time $T$ is finite, the frequency domain will be discretized in steps of $\delta\omega=2\pi/T$. 
In going from the first to the second line, we separated the frequency domain into two disjoint regions: the null-band (NB) $\mB$ and its complement (CNB) $\overline{\mB}$.
As described in the main text (see  Sec.~\ref{subsec:opt-prob}), the NB is defined as the largest (not necessarily connected) subset of frequencies over which the PSD has fractional powers in the NBs $\epsilon_\mu\ll1$, i.e.
\eq{
\frac{1}{P_\mu} \sum_{n\in\mathcal{N}_\mu} S_{\mu,n}\delta\omega = \epsilon_\mu,
}
where $P_\mu=\sum_{n\in\mathcal{N}_\mu\cup\overline{\mathcal{N}}_\mu}
S_{\mu,n}\delta\omega$ is the power along the $\mu$-th channel.
The NB and CNB are normalized and discretized into their discrete versions $\mathcal{N}_{\mu},\,\overline{\mathcal{N}}_{\mu}=\lfloor \mB_{\mu}/\delta\omega \rfloor, \,\lfloor \overline{\mB}_{\mu}/\delta\omega \rfloor$. The sets $\mathcal{N}_{\mu},\,\overline{\mathcal{N}}_{\mu}$ are disjoint subsets of natural numbers satisfying $\mathcal{N}_{\mu}\cup\overline{\mathcal{N}}_{\mu}=[0,...,N-1]$, where $N=T/\delta t$ is the number of time steps for the control functions. 
The time step $\delta t$ depends on hardware limitations.
The discretized FFs $F_{\mu,n}$ are defined as the averages over the frequency windows $[n\delta\omega,(n+1)\delta\omega)$ of the filter functions $F_{\mu,n}=\delta\omega^{-1}\int_{n\delta\omega}^{(n+1)\delta\omega}F_{\mu}(\omega)d\omega$, for $n=0..N-1$. A similar statement can be made for the discrete PSD $S_{\mu,n}$.

Interpreting the sums (or integrals in the continuous case) over frequencies as 1-norms $||\cdot||_1=\sum_{n\in X} |\cdot|$ for some subset of integers $X$, we can use H\"{o}lder's inequality $||fg||_1\leq||f||_2||g||_2$ \cite{hardy1988inequalities} to bound these expressions, where the 2-norm is $||\cdot||_2=\sqrt{\sum_{n\in X} |\cdot|^2}$.
Additionally, since in practice these are finite dimensional spaces, we know that the sum of a positive sequence $f_n$ will satisfy the triangle inequality $\sum_{n} f^2_n\leq (\sum_{n} f_n)^2$. Consequently, we have 
\eq{
\label{eq:norms-bounds}
||f||_2=\sqrt{\sum_{n\in X} f^2_n \delta\omega} 
&\leq \sum_{n\in X} f_n \delta\omega^{1/2} = \delta\omega^{-1/2} ||f||_1
}
where the summation is performed over the discretized frequency regions $X=\{\mathcal{N}_{\mu}, \overline{\mathcal{N}}_{\mu}\}$.
We can use this inequality to bound the sums over $\{\mathcal{N}_{\mu}, \overline{\mathcal{N}}_{\mu}\}$ as follows:
\eq{
\sum_{n\in\mathcal{N}_{\mu}} S_{\mu,n} F_{\mu,n} \delta\omega &\leq \delta\omega \sum_{n\in\mathcal{N}_\mu} S_{\mu,n} \sum_{m\in\mathcal{N}_\mu} F_{\mu,m}\nonumber\\
&\leq  \delta\omega^{-1} \epsilon_{\mu} P_{\mu} T.
\label{eq:bound-nb}
}
Additionally, we have used $\sum_{n=0}^{N-1}F_{\mu,n}\delta\omega=T$, i.e., the integral of the FF over all frequencies is equivalent to the total time.
Note that in the optimal case, where all of the spectral weight of the PSD is in the CNB ($\epsilon_\mu=0$)
these terms in Eq.~(\ref{eq:bound-nb}) all converge to zero.

Similarly, using Eq.~(\ref{eq:norms-bounds}), we bound the integral over the CNB
\eq{
\sum_{n\in\overline{\mathcal{N}}_\mu} S_{\mu,n} F_{\mu,n} \delta\omega &\leq \delta\omega \sum_{n\in\overline{\mathcal{N}}_\mu} S_{\mu,n} \sum_{m\in\overline{\mathcal{N}}_\mu} F_{\mu,m}\nonumber\\
&\leq  \delta\omega^{-1} (1-\epsilon_\mu) P_\mu T \, \Gamma_\mu(T),
\label{eq:bound-cnb}
}
with $\Gamma_\mu(T)=T^{-1}\sum_{n\in\overline{\mathcal{N}}_\mu}F_{\mu,n}\delta\omega$, where the factor of $T^{-1}$ is added explicitly to keep the functions $\Gamma_\mu$ dimensionless.
Combining the bounds from Eqs.~(\ref{eq:bound-nb}) and (\ref{eq:bound-cnb}) and requesting the same level of noise spectral concentration along all axes, i.e. $\epsilon_\mu=\epsilon\forall \mu$, we find
\eq{
2\pi\chi(T) &= \sum_{\mu=x,y,z} \left( \sum_{n\in\overline{\mathcal{N}}_\mu} + \sum_{n\in\mathcal{N}_\mu} \right) S_{\mu,n} F_{\mu,n} \delta\omega \nonumber\\
&\leq \frac{T}{\delta\omega} \sum_{\mu=x,y,z} P_\mu [ (1-\epsilon) \Gamma_\mu(T) +  \epsilon] \nonumber\\
&= \frac{PT}{\delta\omega}\, \Gamma(T) + O(\epsilon).
}
Here, we have used $P=\sum_{\mu=x,y,z} P_\mu$ and introduced
\eq{
\label{eq:objective_function_different_p}
\Gamma(T) = \frac{\delta\omega}{T} \sum_{\mu=x,y,z}p_\mu\sum_{n\in\overline{\mathcal{N}}_\mu}F_{\mu,n},
}
where $p_\mu=P_\mu/P$.
This implies that, to zeroth order in the fractional power, the noise fidelity is bounded by
\eq{
\label{eq:Fgamma-FN}
\mathcal{F}_N(T) &= \frac{1+e^{-\chi(T)}}{2} \geq  \frac{1+e^{- \frac{PT}{\delta\omega} \Gamma(T)}}{2} + O(\epsilon)\nonumber\\
&= \mathcal{F}_\Gamma(T) + O(\epsilon).
}

Finally, the distance will be bounded by
\eq{
\label{eq:fidelity_upper_bound}
\langle D(U_G,U(T))^2\rangle  &\leq 2-\mathcal{F}_G(T) - \mathcal{F}_\Gamma(T) + \mathcal{K}(T) + O(\epsilon),
}
which, as long as $\epsilon$ can be kept small, justifies $\Gamma(T)$ in Eq.~(\ref{eq:objective_function}) as our choice of objective function.
In practice, the gate fidelity $\mF_G(T)$ can be kept as close to 1 as desired by setting it as a constraint in a constrained optimization using an optimizer such as SLSQP.
In the main text, we show in simulation how this assumption yields good noise filtering controls.

In the bound above, $\mF_\Gamma(T)$ is dependent upon the fractional noise power within the CNB, which can place additional requirements on noise characterization protocols. The spectral leakage in Eq.~(\ref{eq:objective_function_different_p}) is composed of a sum of terms, each weighted by the power weights in the $\mu$-th direction $p_\mu$. While refined estimates of noise power may require more QNS resources than those required to determine spectral cutoffs, even rough estimates of noise PSDs can provide sufficient information to identify dominant noise channels.
We denote these estimates as $\overline{p}_\mu$ within the F-GRAFS objective function
\eq{
\label{eq:objective_function}
\Gamma(T) &= \frac{\delta\omega}{T} \sum_{\mu=x,y,z}\overline{p}_\mu\sum_{n\in\overline{\mathcal{N}}_\mu}F_{\mu,n} \\
&= \frac{1}{3T} \sum_{\mu=x,y,z}\overline{p}_\mu\int_{\overline{\mathcal{B}}_\mu}F_{\mu}(\omega)d\omega,
}
where the final expression denotes the continuous frequency representation.

Two distinct scenarios arise from this definition. 
In the case when such estimates reveal a highly asymmetric noise scenario, one can approximate the spectral leakage by the FF corresponding to the dominant noise source. We denote this configuration as the single-axis noise case. In the main text we discuss the case where noise is dominant along the $z$ direction, i.e., $\overline{p}_z\approx1$ and $\overline{p}_x\approx\overline{p}_y\approx0$.
On the other hand, the symmetric case can be considered, where the noise power along all channels is nearly equivalent, i.e., $\overline{p}_\mu\approx1/3$ for all $\mu=x,y,z$.
In the main text we refer to this noise configuration as the multi-axis noise case.
In practice, the fractional noise estimates will lie in between 0 and 1, with the condition that $\sum_\mu \overline{p}_\mu=1$.

Lastly, one could consider choosing an alternative objective function when a reliable estimate of the total power of the noise is available. 
When this is the case, it is possible to use the combined fidelity
\eq{
\label{eq:combined-fidelity}
\Phi(T) = \mathcal{F}_G(T) + \mathcal{F}_\Gamma(T)
}
as an objective function instead of $\Gamma(T)$.
The gradient can be computed following the same steps as described in the main text Sec.~\ref{sec:F-GRAFS-gradients} and using the chain rule.
The combined fidelity allows us to optimize without constraints, for example utilizing an optimizer like L-BFGS-B. 
See Appendix~\ref{sec:optimizers} for further information on the choice of optimization methods.

\section{Optimizer comparison} 
\label{sec:optimizers}

The F-GRAFS optimization algorithm described in Sec.~\ref{sec:F-GRAFS-gradients} can be executed using different gradient descent methods. 
In this section, we compare the following algorithms: SLSQP, L-BFGS-B, trust-region constrained (TRC) and Nelder-Mead (NM). We study the performance of each algorithm as a function of wall time. Algorithmic performance is examined using $\Phi(T)$ or $\Gamma(T)$ as an objective function. The latter being only applicable for SLSQP and TRC. We denote those using $\Phi(T)$ as unconstrained, e.g., unconstrained SLSQP is denoted by U-SLSQP. Similarly, algorithms utilzing $\Gamma(T)$ are defined as constrained, e.g., constrained SLSQP is designated by C-SLSQP.

\begin{figure*}[t]
\centering
\includegraphics[width=.87\textwidth]{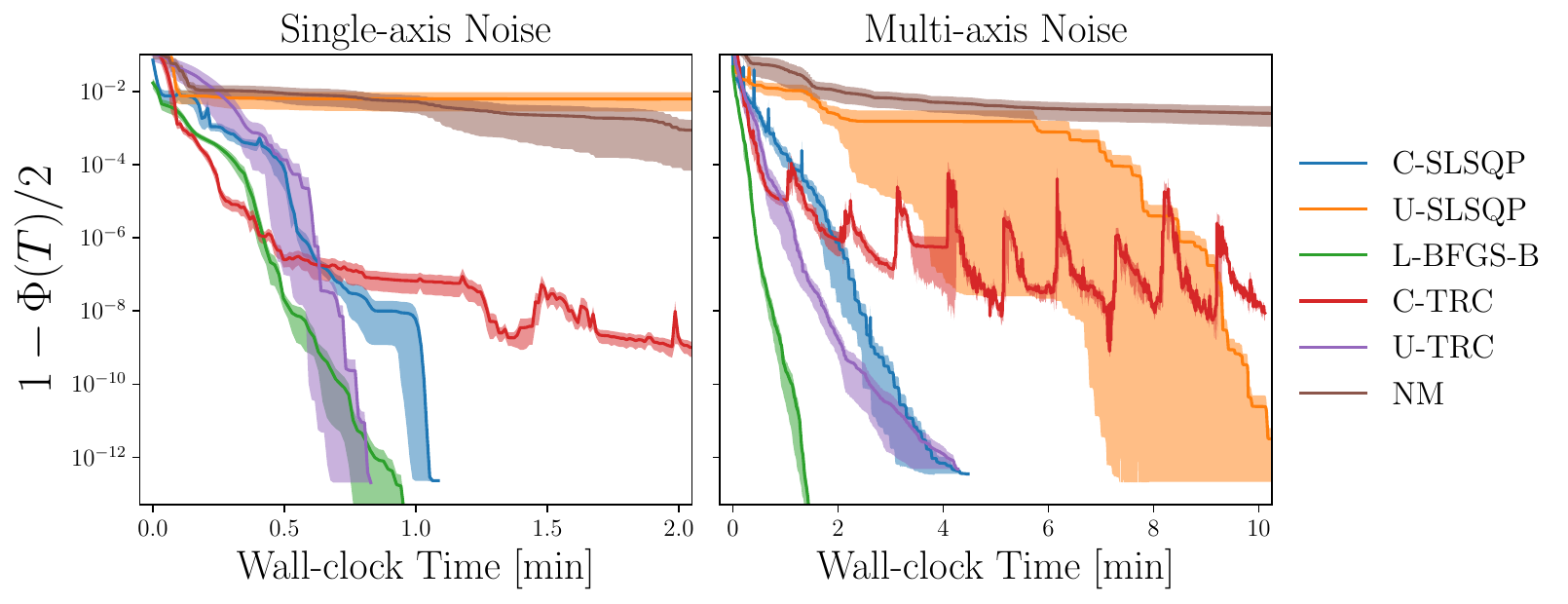}
\caption{F-GRAFS performance of different optimization methods as a function of optimizing time, averaged over different implemented high-pass gates under single-axis (left) and multi-axis (right) noise cases.
In the former, each curve represents the average over eight $X$-rotations uniformly distributed in $[0,2\pi)$, and the Clifford+$T$ gates in the latter.
The shaded regions represent the standard error over these averages.
In both cases, the high-frequency noise cutoff was set to $\omega_H=0.08\times2\pi/\delta t$.
The single-axis cases undergo only dephasing ($z$ axis) noise and $x$ axis control, while the multi-axis noise see $x,y,z$ noise and $x,y$ control.
For all methods considered, optimizations using $\Phi(T)$ were performed (U-SLSQP, L-BFGS-B,U-TRC,NM).
For the methods that allow for constrained optimizations, additional optimizations using $\Gamma(T)$ as objective function (C-SLSQP, C-TRC labels) and treating the gate fidelity as a constraint were studied.
In the single-axis case, it is clear that the methods achieving the best performance are C-SLSQP, L-BFGS-B and U-TRC methods. 
In the multi-axis case, L-BFGS-B presents the best performance. All runs were performed with an Intel(R) Core(TM) i7-10510U CPU @ 1.80GHz 2.30 GHz processor.
}
\label{fig:optimizers}
\end{figure*}

In Fig.~\ref{fig:optimizers}, we present a summary of our results for single-axis (left) and multi-axis (right) noise and control. Here, we see that the methods C-SLSQP (blue), L-BFGS-B (green), and U-TRC (purple) are the only algorithms that consistently achieve the desired levels of objective function reduction within the expected timeframe.
In the single-axis case, L-BFGS-B and U-TRC are indistinguishable within errorbars, and present only a slight advantage with respect to C-SLSQP. 
In the multi-axis case, L-BFGS-B is the fastest method by approximately a factor of 3, while U-TRC and C-SLSQP are indistinguishable within errorbars. 
From this, it can be concluded that the fastest method to optimize $\Phi(T)$ is L-BFGS-B, and should be used if knowledge of the total noise power is available.
On the other hand, C-SLSQP performs the best for the constrained optimization of $\Gamma(T)$ subject to $\mF_G(T)>1-\epsilon_G$, for some $\epsilon_G\ll1$.

\section{Initial conditions and control ansatz} 
\label{sec:initial_conditions}
\subsection{Single-axis: Constant Drive}
\label{sec:SA_IC}

The convenience of utilizing CD as initial condition for the optimization can be seen by analytically computing the associated FF.
For the present analysis, we consider the scenario with single-axis control along $x$ with dephasing noise along $z$.
In this case, the FF takes the form
\eq{
\label{eq:single-axis-FF}
F(\omega,T)=\left|\int_0^T \cos\Theta(t)e^{-i\omega t}dt\right|^2+\left|\int_0^T \sin\Theta(t)e^{-i\omega t}dt\right|^2,
}
where $T$ is the total time.
CD control is achieved by setting $\Omega(t)=\Omega_0$, from where we obtain $\Theta(t) = \int_0^t \Omega(s)ds= \Omega_0 t$. 
Using $\Theta(t)$ in Eq.~(\ref{eq:single-axis-FF}), it is possible to compute the integral explicitly, obtaining
\eqs{
\label{eq:single-axis-FF-CD}
&\frac{F(\omega,T)}{T^2} =  \frac{2}{T^2\left(\omega ^2-\Omega_0 ^2\right)^2}\Big[\left(\omega ^2+\Omega_0 ^2\right) \times \\
&\quad \times (1-\cos (\omega T) \cos (\Omega_0 T)) -2 \omega  \Omega_0  \sin (\omega T) \sin (\Omega_0 T)\Big] \\
&\quad \overset{T\rightarrow\infty}{\longrightarrow}
\frac{1}{2} \left(\delta(\omega+\Omega_0)+\delta(\omega-\Omega_0)\right),
}
where in the second line we consider the infinite $T$ limit.
It is straightforward to see that in this limit, the FF converges to delta functions centered around $\pm \Omega_0$, normalized such that $\delta(0)=1$. 
From Eq.~(\ref{eq:single-axis-FF-CD}), it follows that Eq.~(\ref{eq:overlap_general}) leads to an overlap $\chi(T) = TS(\Omega_0)$, where we used the fact that semiclassical noise PSDs are even around $\omega=0$. 
In order to minimize the overlap, the driving frequency should be tuned to the minimum of the PSD, i.e., $\Omega_0=\mathrm{argmin}_{\omega}S(\omega)$.
It is worth studying the case of monotonically decreasing $S(\omega)$, e.g., $1/f$ noise, where no global minima exist. In order to reduce the overlap between the FF and the PSD, $\Omega_0$ should be chosen as large as allowed by hardware, provided that it does not violate additional constraints on the control.

An additional argument in favor of the CD control ansatz comes from noting that CD is the solution with minimum power for a given rotation angle $\Theta(T) = \int_0^T \Omega(t) dt=\Omega_0 T$.
Suppose another control function $\tilde{\Omega}(t)$ produces the same rotation angle, i.e., $\int_0^T \tilde{\Omega}(t) dt = \Theta(T)$, then the power of this new control is
\begin{eqnarray}
\int_0^T \tilde{\Omega}(t)^2 dt &=& 
\int_0^T [ \Omega_0 + (\tilde{\Omega}(t)-\Omega_0)]^2 dt  = \nonumber \\
&\stackrel{(1)}{=}& \int_0^T \Omega_0^2 dt +
\int_0^T [\tilde{\Omega}(t)-\Omega_0]^2 dt \nonumber \\
&\stackrel{(2)}{\geq}& \int_0^T \Omega(t)^2 dt.
\end{eqnarray}
In (1) we used the assumption that the area of the difference is zero $\int_0^T[ \tilde{\Omega}(t)-\Omega_0]dt =0$ to cancel the crossterm appearing from expanding the quadratic function in the integrand in the first line. 
Inequality (2) comes from the fact that $\int_0^T [\tilde{\Omega}(t)-\Omega_0]^2 dt\geq0$, and implies that the power of $\tilde{\Omega}(t)$ is larger than that of CD control.

\begin{figure}[h]
\centering
\includegraphics[width=.45\textwidth]{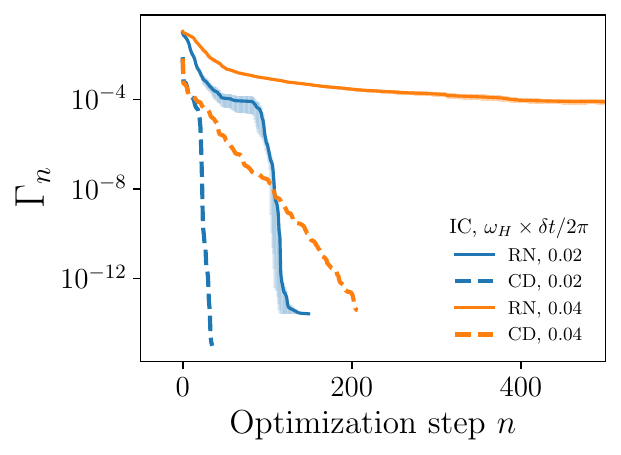}
\caption{Comparison of objective function $\Gamma(T)$ as function of the optimization step $n$, which we note $\Gamma_n$, between random initial conditions (RN, solid lines) averaged over 20 different realizations, and constant drive (CD, dashed lines) initial conditions for single axis noise and control. The shaded regions represent the standard error at each step of the randomly initialized optimizations.
For the blue lines, the CNB is $\overline{\mathcal{B}}=[0,0.02)\times2\pi/\delta t$, while for the orange lines possess a larger CNB of  $\overline{\mathcal{B}}=[0,0.04)\times2\pi/\delta t$. 
The tolerance on the objective function gradient is set to $|\Gamma_{n+1}-\Gamma_{n}|=10^{-15}$. 
For these optimizations, L-BFGS-B was used with Eq.~(\ref{eq:combined-fidelity}) as objective function.
Note how for the lower noise cutoff, the optimization using CD as initial condition improves exponentially until it stops before the step 50 at about $\Gamma_n=10^{-14}$, while the optimizations using random initial conditions averaged over ten realizations converge to similarly good solutions but taking about 150 steps.
For the higher noise cutoff, the random initial conditions are not able to find good noise filtering solutions in average.
This shows how CD can correct the instability of choosing random initial conditions, adding robustness, while at the same time improve the running time, since it not only satisfies the tolerance earlier, but it also does not require selecting from multiple realizations.
}
\label{fig:RNvsCD}
\end{figure}

Lastly, we can see in Fig. \ref{fig:RNvsCD} that using CD (dashed lines) as initial condition of F-GRAFS provides a qualitative advantage over random initialization (RN, solid lines). 
Each curve represent the values of the objective function $\Gamma(T)$ as a function of the optimization steps. 
The optimizations produce high-pass filters implementing identity gates, with high-frequency cutoffs of $\omega_H=0.02\times2\pi/\delta t$ (blue) and $0.04\times2\pi/\delta t$ (orange).
While CD initilization is run once, random initializations are averaged over 20 different realizations. 
For the lower frequency noise, we see that CD finds a solution with $\Gamma(T)<10^{-14}$ in approximately 30 steps. Random initial conditions, on the other hand, take about 150 steps.
For the higher cutoff case, CD initialization reaches the desired solution in about 200 steps, while RN is not capable of converging within 1000 steps.
This example highlights the importance of using CD in minimizing the computational cost of the F-GRAFS optimization. 

\subsection{Multi-axis noise and control}
\label{sec:MA_IC}
\subsubsection{Derivation of Initial Conditions}
In the multi-axis control and noise scenario, the quality of F-GRAFS solutions varies extensively when employing random initial conditions. In order to get consistent cancellation over the CNB and therefore improve algorithmic stability, it is necessary to narrow down the search space.
Although CD control is an effective solution for the single axis case, any combination of CD in the $x$ and $y$ axes will lead to FFs with non-zero DC contributions. 
The multi-axis CD condition is given by  $\vOmega(t)=(\Omega_x,\Omega_y,0)=\Omega_0 (\cos\phi_0, \sin\phi_0,0)$, where the amplitude $\Omega_0$ and phase $\phi_0$ correspond to the control representation in polar coordinates in the $xy$ plane.
The presence of the DC component can be more easily interpreted in the case where noise is symmetric along $x$ and $y$ axes, implying $S_x(\omega)=S_y(\omega)$.  
In this case, the two resulting FFs are
\eqs{
&F_z(\omega,T) = \left| \int_0^T \cos(\Omega_0 t)e^{-i\omega t} dt \right|^2 + \left| \int_0^T \sin(\Omega_0 t)e^{-i\omega t}dt \right|^2 \\
&F_{x}(\omega,T)+F_{y}(\omega,T) = F_z(\omega,T)+T^2\sinc^2\pr{\omega T/2}.
}
Here, $F_z(\omega,T)$ is the same as in the single-axis case, but the longitudinal FFs $F_{x}(\omega,T),F_{y}(\omega,T)$ have an extra term involving a $\sinc(x)$ function. 
Note that this combination of FFs is independent of the angle of $\phi_0$ and only depend on the amplitude $\Omega_0$.
The reason behind this is that multi-axis CD control, although acting on both axes simultaneously, is still a single axis control along the rotated direction given by $\phi_0$.
As shown in Sec.~(\ref{sec:SA_IC}), it is possible to choose the value of $\Omega_0$ such that $F_z(\omega,T)$ does not present relevant low frequency contributions. 
On the other hand, the $\sinc^2\pr{\omega T/2}$ term in $F_x(\omega,T)+F_y(\omega,T)$ is concentrated around $\omega=0$ and is independent of the control. Consequently, there is no choice of multi-axis CD parameters that provides DC filtering along $x$ and $y$ axis simultaneously. This DC component makes multi-axis CD alone a poor choice for initial conditions when noise exists along multiple directions.

Another single-axis solution for low frequency noise that provides an intuitive way of shaping the FFs is oscillating control $\Omega(t)=\Omega_0\cos(\lambda t)$, where $\lambda$ can typically be chosen as the high frequency cutoff $\omega_H$.
These controls produce FFs that can be thought of as a frequency comb in $\lambda$ and modulated by Bessel functions; more specifically $F(\omega)=T^2\sum_{k\in\mathbb{Z}}\delta(\omega-k\lambda)J_k(\Omega_0/\lambda)$ \red{\cite{barr2022qcns}}. In order to cancel DC noise contributions, the control amplitude is chosen as $\Omega_0=\lambda x_0$, where $x_0$ is the first zero of the zeroth order Bessel function.

Motivated by oscillating and CD single-axis controls, as well as work with Walsh synthesized filters \cite{Ball2015Walsh}, we analyze the FFs obtained from CD control along one axis, and an oscillating square wave on the other axis. This is, we study the Hamiltonian in Eq.~(\ref{eq:H_Ctrl}), with
\eq{
\Omega_x(t) &= \Omega_{CD} = \Omega_0 \cos(\phi_0)\\
\Omega_y(t) &= \Omega_{SW} s_\lambda(t) = \Omega_0 \sin (\phi_0) s_\lambda(t),
}
where $s_\lambda(t)$ is a square wave with unit amplitude and frequency $\lambda$, which defines the control period $T_c=2\pi/\lambda$.
The polar control coordinates are $\Omega_0=\sqrt{\Omega_{CD}^2+\Omega_{SW}^2}$, and $\cos\phi_0=\Omega_{CD}/\Omega_0, \sin\phi_0=\Omega_{SW}/\Omega_0$, which will be the representation used in the following discussion.
In these coordinates, the control can be interpreted as having constant amplitude $\Omega_0$, and alternating phase $\phi(t)=\pm\phi_0$, depending on whether $m T_c<t<(m+\tfrac{1}{2})T_c$ or $(m+\tfrac{1}{2})T_c<t<(m+1)T_c$, for $m=0,1,..,M-1$. Here, $M=T/T_c$ denotes the number of control periods that fit in the full control time duration.
Note that in terms of the control frequency, $M=\lambda T/2\pi = \lambda/\delta\omega$, meaning that $M$ is the discrete, normalized control frequency.
Hence, in what follows we will refer to $M$ as a control parameter, rather than $\lambda$.
The goal is, for fixed total time $T$, to find optimal parameters $\Omega_0, \phi_0, M$ that minimize the FFs over a given CNB. 
We show next that it is possible to derive optimal values for control parameters that produce high-pass filters in all three axes.

The argument is as follows: first, we construct the noise Hamiltonian in the toggling frame with respect to the control; then we use this to extract the control matrix $R(t)$ whose Fourier transform $R(\omega)$ is directly related to the FFs through Eq.~(\ref{eq:FF_R_general}).
Lastly, we show that by rewriting $R(\omega)$ in a convenient way, it is possible to cancel all elements of the frequency space control matrix for some frequencies $\omega$, and hence cancel the FFs within the CNBs.

In general, the time evolution operator will be hard to compute analytically. However, in the piece-wise constant control case, it is possible to arrive at a closed expression.
Given the noise Hamiltonian in Eq~(\ref{eq:H-noise}), the toggling frame Hamiltonian $\tilde{H}(t)$ is computed by 
$\tilde{H}(t)= U_C(t)^\dagger H_N(t) U_C(t)$,
where $U_C(t)$ is the control time propagator.
Due to the periodicity of the control, the control propagator takes the form
\begin{widetext}
\eq{
U_C(t)&=\begin{cases}
e^{-i \hn^+\cdot\vsigma \frac{\Omega_0}{2} (t-m T_c) } \left(
e^{-i \hn^-\cdot\vsigma \frac{\Omega_0}{2} \frac{T_c}{2} }
e^{-i \hn^+\cdot\vsigma \frac{\Omega_0}{2} \frac{T_c}{2} }
\right)^m, 
& m T_c<t<(m+\frac{1}{2})T_c \\
e^{-i \hn^-\cdot\vsigma \frac{\Omega_0}{2} (t- (m+\frac{1}{2}) T_c) }
e^{-i \hn^+\cdot\vsigma \frac{\Omega_0}{2} \frac{T_c}{2} }
\left(
e^{-i \hn^-\cdot\vsigma \frac{\Omega_0}{2} \frac{T_c}{2} }
e^{-i \hn^+\cdot\vsigma \frac{\Omega_0}{2} \frac{T_c}{2} }
\right)^m, 
& (m+\frac{1}{2}) T_c<t<(m+1)T_c \\
\end{cases}
}
\end{widetext}
for $m=0,..,M-1$, where $\hn^{\pm}=(\Omega_{CD},\pm\Omega_{SW},0)/\Omega_0 =(\cos\phi_0,\sin\pm\phi_0,0)$. 

The toggling frame Hamiltonian in the interval $0<t<T_c/2$ is
\begin{eqnarray}
\tilde{H}(t)
&=& e^{i \hn^+\cdot\vsigma \frac{\Omega_0}{2} t } \vbeta(t)\cdot\vsigma e^{-i \hn^+\cdot\vsigma \frac{\Omega_0}{2} t } \nonumber\\
&=& \vsigma \cdot  \mathcal{R}_{\Omega_0,\phi_0}(t) \cdot \vbeta(t) ,
\end{eqnarray}
where we have used the Rodrigues' rotation formula \cite{rodrigues_formula}. Additionally, we define the half-period rotation matrix
\begin{widetext}
\eq{
\mathcal{R}_{\Omega_0,\phi_0}(t)=
\begin{bmatrix} 
\sin^2\phi_0 \cos(\Omega_0 t) + \cos^2\phi_0 & \sin\phi_0\cos\phi_0(1-\cos(\Omega_0 t)) & -\sin\phi_0\sin(\Omega_0 t)\\
\sin\phi_0\cos\phi_0(1-\cos(\Omega_0 t)) & \cos^2\phi_0 \cos(\Omega_0 t) + \sin^2\phi_0 & \cos\phi_0\sin(\Omega_0 t) \\
\sin\phi_0\sin(\Omega_0 t) & - \cos\phi_0\sin(\Omega_0 t) & \cos(\Omega_0 t)
\end{bmatrix}.
}
\end{widetext}
which captures the evolution induced by the control within a constant section of the control functions.
The matrix $\RR_{\Omega_0,\phi_0}(t)$ will be the building block of the full evolution.
Note that it can be conveniently written, using Euler decomposition, in terms of basic rotations as $\RR_{\Omega_0,\phi_0}(t)=R_z(\phi_0)R_x(-\Omega_0 t)R_z(-\phi_0)$.
Next, we compute the toggling frame Hamiltonian for the interval $T_c/2<t<T_c$,
\eq{
\tilde{H}(t)
&= e^{i \hn^+\cdot\vsigma \frac{\Omega_0}{2} \frac{T_c}{2} } 
e^{i \hn^-\cdot\vsigma \frac{\Omega_0}{2} (t-\frac{T_c}{2}) } \Big[
\vbeta(t)
\cdot\vsigma 
\Big] \times \nonumber\\
&\qquad \times
e^{-i \hn^-\cdot\vsigma \frac{\Omega_0}{2} (t-\frac{T_c}{2}) } 
e^{-i \hn^+\cdot\vsigma \frac{\Omega_0}{2} \frac{T_c}{2} } \nonumber\\
&= e^{i \hn^+\cdot\vsigma \frac{\Omega_0}{2} \frac{T_c}{2} }
\left[
\mathcal{R}_{\Omega_0,-\phi_0}(t-\tfrac{T_c}{2}) \cdot \vbeta(t)
\right]
\cdot\vsigma 
e^{-i \hn^+\cdot\vsigma \frac{\Omega_0}{2} \frac{T_c}{2} } \nonumber\\
&=  \vsigma \cdot \left[
\RR_{\Omega_0,\phi_0}(\tfrac{T_c}{2})\cdot \RR_{\Omega_0,-\phi_0}(t-\tfrac{T_c}{2}) \cdot \vbeta(t) \right].
}
From here it becomes clear that the action of the control consists of successive applications of the constant section rotation matrix, with variations in the parameters controlling the rotation. 

The toggling frame Hamiltonian can then be constructed iteratively, yielding
\begin{widetext}
\eq{
\tH(t) &= \begin{cases}
\vsigma \cdot \Large\{ \left[
\RR_{\Omega_0,\phi_0}(\tfrac{T_c}{2})\cdot \RR_{\Omega_0,-\phi_0}(\tfrac{T_c}{2})
\right]^m \cdot 
\RR_{\Omega_0,\phi_0}(t-m T_c) \cdot \vbeta(t) \Large\}&, m T_c<t<(m+\frac{1}{2})T_c \nonumber\\
\vsigma \cdot \Large\{ \left[
\RR_{\Omega_0,\phi_0}(\tfrac{T_c}{2})\cdot \RR_{\Omega_0,-\phi_0}(\tfrac{T_c}{2})
\right]^m \cdot 
\mathcal{R}_{\Omega_0,\phi_0}(\tfrac{T_c}{2})
\cdot \mathcal{R}_{\Omega_0,-\phi_0}(t-(m+\tfrac{1}{2}) T_c) \cdot \vbeta(t) \Large\}&, (m+\frac{1}{2}) T_c<t<(m+1)T_c \\
\end{cases}\\
&= \vsigma\cdot R(t) \cdot \vbeta(t).
}

\end{widetext}

Since the FFs depend on the Fourier transforms of the control matrix via Eq.~(\ref{eq:FF_R_general}), we compute

\begin{widetext}

\eq{
R(\omega) &=\int_0^T R(t) e^{-i \omega t} dt = 
\sum_{m=0}^{M-1} \int_{mT_c}^{(m+1)T_c} R(t) e^{-i \omega t} dt \nonumber\\
&= \sum_{m=0}^{M-1} \int_{mT_c}^{(m+1/2)T_c} \left[
\RR_{\Omega_0,\phi_0}(\tfrac{T_c}{2})\cdot \RR_{\Omega_0,-\phi_0}(\tfrac{T_c}{2})
\right]^m \cdot 
\mathcal{R}_{\Omega_0,\phi_0}(t-m T_c) e^{-i \omega t} dt + \nonumber \\
&\hspace{.6cm}+ \int_{(m+1/2)T_c}^{(m+1)T_c} \left[
\RR_{\Omega_0,\phi_0}(\tfrac{T_c}{2})\cdot \RR_{\Omega_0,-\phi_0}(\tfrac{T_c}{2})
\right]^m \cdot 
\RR_{\Omega_0,\phi_0}(\tfrac{T_c}{2})
\cdot \RR_{\Omega_0,-\phi_0}(t-(m+\tfrac{1}{2}) T_c) e^{-i \omega t} dt
\nonumber\\
&= \sum_{m=0}^{M-1} \left[
\RR_{\Omega_0,\phi_0}(\tfrac{T_c}{2})\cdot \RR_{\Omega_0,-\phi_0}(\tfrac{T_c}{2})
\right]^m \cdot 
\Big[ 
\int_{mT_c}^{(m+1/2)T_c} \mathcal{R}_{\Omega_0,\phi_0}(t-m T_c) e^{-i \omega t} dt + \nonumber \\
&\hspace{5cm}
\int_{(m+1/2)T_c}^{(m+1)T_c} 
\RR_{\Omega_0,\phi_0}(\tfrac{T_c}{2})
\cdot \RR_{\Omega_0,-\phi_0}(t-(m+\tfrac{1}{2}) T_c) e^{-i \omega t} dt
\Big] \nonumber\\
&= \Big\{\sum_{m=0}^{M-1} \left[
\RR_{\Omega_0,\phi_0}(\tfrac{T_c}{2})\cdot \RR_{\Omega_0,-\phi_0}(\tfrac{T_c}{2})
 e^{-i \omega T_c}
\right]^m \Big\} \cdot \nonumber \\
&\hspace{2cm}\cdot \Big[ 
\int_{0}^{T_c/2} \mathcal{R}_{\Omega_0,\phi_0}(t) e^{-i \omega t} dt + 
e^{-i \omega \tfrac{T_c}{2}} \RR_{\Omega_0,\phi_0}(\tfrac{T_c}{2}) \cdot 
\int_{0}^{T_c/2} \RR_{\Omega_0,-\phi_0}(t) e^{-i \omega t} dt
\Big] \nonumber\\
&= \Big\{\sum_{m=0}^{M-1} \left[
\RR_{\Omega_0,\phi_0}(\tfrac{T}{2M})\cdot \RR_{\Omega_0,-\phi_0}(\tfrac{T}{2M})
 e^{-i \tfrac{\omega T}{M}}
\right]^m \Big\} \cdot \Big[ 
\Phi_{\Omega_0,}(\omega) + 
e^{-i \tfrac{\omega T}{2M}} \RR_{\Omega_0,}(\tfrac{T}{2M}) \cdot 
\Phi_{\Omega_0,-}(\omega)
\Big] \nonumber\\
&= R_\Sigma(\omega) \cdot R_\Phi(\omega).
}
\end{widetext}
Here we defined 
\eq{
R_\Sigma(\omega) &= \sum_{m=0}^{M-1} \left[
\RR_{\Omega_0,\phi_0}(\tfrac{T}{2M})\cdot \RR_{\Omega_0,-\phi_0}(\tfrac{T}{2M})
 e^{-i \tfrac{\omega T}{M}}
\right]^m, \\
R_\Phi(\omega) &= 
\Phi_{\Omega_0,\phi_0}(\omega) + 
e^{-i \tfrac{\omega T}{2M}} \RR_{\Omega_0,\phi_0}(\tfrac{T}{2M}) \cdot 
\Phi_{\Omega_0,-\phi_0}(\omega),
}
where the Fourier transform matrix is
\eq{
\Phi_{\Omega_0,\phi_0}(\omega) &= \int_{0}^{\tfrac{T}{2M}} \mathcal{R}_{\Omega_0,\phi_0}(t) e^{-i \omega t} dt.
}
Note that $R(\omega)$ is now a matrix product of the matrices $R_\Sigma(\omega)$ and $R_\Phi(\omega)$. 
The first one is a geometric sum over rotation matrices multiplied by a complex exponential.
The second one consists essentially of the Fourier transforms over a single control period.
We have thus arrived at an expression for the frequency domain representation of the control matrix in terms of the constant-section control matrix $\RR_{\Omega_0,\phi_0}(t)$ and its Fourier transform.

The matrix $R_\Sigma(\omega)$ can be summed explicitly by diagonalizing $\RR_{\Omega_0,\phi_0}(\tfrac{T}{2M})\cdot \RR_{\Omega_0,-\phi_0}(\tfrac{T}{2M})$, which being a combination of rotation matrices is another rotation matrix.
The diagonal form of the rotation matrix resulting from the product is $\mathcal{D}_{\Omega_0,\phi_0}(\tfrac{T}{2M})=\mrm{diag}(1,e^{i\theta},e^{-i\theta})$, where $\theta$ depends on $\Omega_0,\phi_0,T,M$ and is the rotation angle performed by the combined rotation. 
The relationship between $\theta$ and the control parameters can be found using the following property: for any rotation matrix $R$ with rotation angle $\theta$, its trace is $\mrm{Tr}R = (1+\cos\theta)/2$.
Additionally, we can find the vector $\vec{u}$ parallel to the rotation axis satisfying $R\vec{u}=\vec{u}$ by using the formula $[\vec{u}]_\times = R-R^T$, where $[\cdot]_\times$ denotes the cross product matrix for the vector $\vec{u}$.
Applying $R=\RR_{\Omega_0,\phi_0}(\tfrac{T}{2M})\cdot \RR_{\Omega_0,-\phi_0}(\tfrac{T}{2M})$, these equations yield expressions for the rotation angle
\eqs{
\label{eq:cos_theta_ful}
\cos\theta &= 
\cos^4(\phi_0 ) \cos (\tfrac{\Omega_0 T}{M} ) \\
&- \frac{1}{2} \sin ^2(\phi_0 ) \left(-8 \cos^2(\phi_0 ) \cos ( \tfrac{\Omega_0 T}{2M})+3 \cos (2 \phi_0 )+1\right),
}
and rotation axis
\eq{
\vec{u}_{\Omega_0,\phi_0} = (1,0,\sin\phi_0\tan(\tfrac{T\Omega_0}{4M})),
}
in terms of the control parameters $\Omega_0,\phi_0,T,M$.

In the basis where $\RR_{\Omega_0,\phi_0}(\tfrac{T}{2M})\cdot \RR_{\Omega_0,-\phi_0}(\tfrac{T}{2M})$ is diagonal, $R_\Sigma(\omega)=\sum_{m=0}^{M-1}  \left[\mathcal{D}_{\Omega_0,\phi_0}(\tfrac{T}{2M}) 
e^{-i \omega T/M} \right]^m$, where each diagonal entry is a geometric sum that can be summed explicitly. 
With finite total time $T$, frequency values are discretized in steps of $\delta \omega = 2\pi/T$, taking the continuous frequencies to discrete frequencies $\omega\rightarrow\omega_n=n\delta \omega=n 2 \pi/T$. 
The diagonal $R_\Sigma(\omega)$ then becomes

\begin{eqnarray}
R_\Sigma(\omega=\tfrac{2\pi n}{T})
&=& \sum_{m=0}^{M-1}  \left[
\begin{pmatrix}
1 & 0 & 0 \\
0 & e^{i\theta} & 0 \\
0 & 0 & e^{-i\theta}
\end{pmatrix}
 e^{-i \omega T/M} 
\right]^m \nonumber\\
&=& \begin{pmatrix}
\frac{1-e^{-i 2\pi n}}{1-e^{-i 2 \pi n/M}} & 0 & 0 \\
0 &\frac{1-e^{-i(2\pi n-\theta M)}}{1-e^{-i(2\pi n/M-\theta)}} & 0 \\
0 & 0 &\frac{1-e^{-i(2\pi n+\theta M)}}{1-e^{-i(2\pi n/M+\theta)}}
\end{pmatrix}.\nonumber\\
\label{eq:diagonal_R_Sigma}
\end{eqnarray}

In general, we aim to cancel this matrix as much as possible over the range of frequencies corresponding to the CNB, which is equivalent to reducing its rank.
Imposing the condition that $\theta M=2\pi \ell, \ell\in\mathbb{Z}$ ensures that this matrix has rank 1 for $n=0$, the value corresponding to the lowest frequency noise contribution $\omega=0$.
Here, the matrix $R_\Sigma(\omega)$ becomes $\mrm{diag}(M,0,0)$, where the non-zero eigenvalue corresponds to the eigenvector parallel to the axis of rotation. 

The diagonal elements of the matrix $R_\Sigma(\omega)$ will have non-zero values only for those frequencies $\omega_n=n\delta\omega$ with $n\in\{M(\ell\pm\theta/2\pi), M\ell\}$, which for the first few values of $\ell\in\mathbb{Z}$ means $n=0,M\theta/2\pi, M(1-\theta/2\pi)$. 
Thus, by choosing the controls parameters $M,\theta$ from the noise cutoffs as
\eq{
\label{eq:MA-M_theta}
M &= n_z+n_{xy}, \\
\theta &= 2\pi \frac{n_z}{M},
}
with $n_z=\omega_z/\delta\omega$ and $n_{xy}=\omega_{xy}/\delta\omega$, we ensure that $R_\Sigma(\omega)$ (and consequently the FF) will be cancelled for all low frequencies except $\omega\in\{0,\omega_z,\omega_{xy}\}$.
Next, we will show how it is possible to cancel the product between $R_\Sigma(\omega)$ and $R_\Phi(\omega)$  by properly $\Omega_0,\phi_0$.

The zero frequency case is of particular interest, since in most practical applications, low frequency noise will be dominant. Having already shown that it is possible to choose $M,\theta$ such that $R_\Sigma(\omega)$ is of rank 1 for $\omega=0$, we examine $R(\omega=0)$ and impose the condition that it vanishes
\eqs{
R(\omega=0) &= \Big\{\sum_{m=0}^{M-1} \left[
\RR_{\Omega_0,\phi_0}(\tfrac{T}{2M})\cdot \RR_{\Omega_0,-\phi_0}(\tfrac{T}{2M})
\right]^m \Big\} \cdot \\
&\quad \cdot \Big[ 
\Phi_{\Omega_0,\phi_0}(0) + 
\RR_{\Omega_0,\phi_0}(\tfrac{T}{2M}) \cdot 
\Phi_{\Omega_0,-\phi_0}(0)
\Big] \\
&= 0
}
The goal is to find values of $\Omega_0,\phi_0$ for which the matrix product between $R_\Sigma$ and $R_\Phi$ is identically zero. From a Linear Algebra perspective, this means setting the values of $\Omega_0,\phi_0$ such that the image of $R_\Phi$ maps to the kernel of $R_\Sigma$. 
Assuming that the control parameters are set as in Eq.~(\ref{eq:MA-M_theta}),
for $\omega=0$ we have that the kernel of $R_\Sigma$ is the subspace orthogonal to the axis of rotation $\vec{u}_{\Omega_0,\phi_0}$.
In other words, we look for $\Omega_0,\phi_0$ that restrict the image of $R_\Phi(0)$ to this orthogonal subspace. 
We can impose this condition by requesting that the inner product between $\vec{u}_{\Omega_0,\phi_0}$ and $R_\Phi(0)\cdot v$ is 0 for all $v\in\mathbb{R}^3$.
It is enough to show this for some basis vectors.
Choosing the Cartesian basis $v_1=(1,0,0),v_2=(0,1,0),v_3=(0,0,1)\}$, we see that this is automatically satisfied for $v_2,v_3$, i.e., $\langle \vec{u}_{\Omega_0,\phi_0}, R_\Phi(0)\cdot v_2 \rangle = \langle \vec{u}_{\Omega_0,\phi_0}, R_\Phi(0)\cdot v_3 \rangle = 0$.
On the other hand, applying this for $v_1$ yields a non-trivial condition for $\Omega_0,\phi_0$ in terms of $M,T$
\begin{eqnarray}
0 &=& \langle \vec{u}_{\Omega_0,\phi_0}, R_\Phi(0)\cdot v \rangle\\
&=& \frac{8 M \sin ^2(\phi_0 ) \tan \left(\frac{T \Omega_0 }{4 M}\right)+2 T \Omega_0  \cos^2(\phi_0 )}{2 M \Omega_0 }\\
& & \Rightarrow \tfrac{\Omega_0 T}{4M} = -\tan^2(\phi_0)\tan(\tfrac{\Omega_0 T}{4M}).
\label{eq:phi_control}
\end{eqnarray}

Next, from Eq.~(\ref{eq:cos_theta_ful}) we can see that one possible solution of this equation for $\Omega_0,\phi_0$ in terms of the other parameters is
\eq{
\cos(\tfrac{\Omega_0 T}{2M}) = 1 - 2\cos^2(\theta/4)[1+\tan^2\phi_0]
}
which, combined with the previous equation results in
\eq{
\frac{\sin^2(\tfrac{\Omega_0 T}{4M})}{1-\tfrac{\Omega_0 T}{4M}\cot(\tfrac{\Omega_0 T}{4M})} = \cos^2(\theta/4).
}
Positive solutions to this equation in terms of $\tfrac{\Omega_0 T}{4M}$ can be easily found numerically using standard tools for each value of $n_z,n_{xy}$ from where it is straightforward to determine $\Omega_0$.
The value of $\phi_0$ can be set by inverting Eq.~(\ref{eq:phi_control}).
Additionally, it can be shown that with this choice of parameters, $F_{xy}(\omega<\omega_{xy})=0$ and $F_{z}(\omega<\omega_{z})=0$.

To summarize, we use the multi-axis CD and square wave scheme to define a low frequency noise filtering problem. 
We start with the control Hamiltonian [Eq.~(\ref{eq:H_Ctrl})] with $\vOmega(t)=\Omega_0(\cos\phi(t),\sin\phi(t),0)$. 
Here, $\phi(t)=\phi_0 s_\lambda(t)$ is a square wave of amplitude $\phi_0$ and frequency $\lambda$.
The problem is defined as finding control parameters $\Omega_0,\phi_0,M$ that filter noise along all three axes up to high frequency noise cutoffs $\omega_z = n_z\delta\omega, \omega_{xy} = n_{xy}\delta\omega$.
We find analytical solutions to this problem by solving the system of equations
\eq{
\begin{cases}
M &= n_z + n_{xy}, \\
\theta &= 2\pi \frac{n_z}{M}, \\
\frac{\sin^2(\xi)}{1-\xi\cot(\xi)} &= \cos^2(\theta/4), \\
\xi &= \frac{\Omega_0 T}{2M},\\
\tan^2\phi_0 &= -\xi \cot(\xi),
\end{cases}.
\label{eq:multiaxis_initial_condition_problem}
}
The third equation can be easily solved numerically with standard tools such as the \textit{optimize.fsolve} function from the \textit{SciPy} Python library, using the value of $\xi_0=2$ as a seed for $\xi$.
In order to choose between the family of solutions found by solving this equation, it is possible to use an argument of efficiency of resources and choose the smallest $x$, i.e., the value that minimize the control amplitude $\Omega_0$.

\subsubsection{Alternative choices of initial conditions}

To conclude this section, we add that solutions achieving equivalent levels of cancellation were found numerically by using a sinusoidal control rather than a square wave. That is, $\Omega_x(t)=\Omega_{CD}$ and $\Omega_y(t)=\Omega_{SW} \sin(\lambda t)$, for given $\Omega_{CD}, \Omega_{SW}$ values, where $\lambda$ can be set as above from $M=n_z+n_{xy}$.
The disadvantage of this control is that we lack an analytical derivation of the control parameters, since the piece-wise constant assumption of the square wave control used in the previous derivation is no longer valid.
Nevertheless, it is straightforward to perform a numerical exploration to find which values of the $\Omega_{CD},\Omega_{SW}$ parameters yield the desired FF features. For example, by using the F-GRAFS gradients in the functional basis of $\{1,\sin(\lambda t)\}$ one can find that the FFs obtained are equivalent to those described in the previous section.

Lastly, it is also possible to create good solutions using a different ansatz, where the controls are chosen to be a combination of CD and the square wave: $\Omega_x(t)=\Omega_0 (\cos\phi_0+\sin\phi_0 s_\lambda(t))/\sqrt{2}$ and $\Omega_y(t)=\Omega_0 (\cos\phi_0-\sin\phi_0 s_\lambda(t))/\sqrt{2}$.
Interestingly, the values of these controls that produce the desired FFs coincide with the ones obtained from solving Eqs.~(\ref{eq:multiaxis_initial_condition_problem}). 
The reason behind this is that these controls only differ from those described in Eqs.~(\ref{eq:multiaxis_initial_condition_problem}) by a $\pi/4$-rotation about the $z$-axis.
The benefit of choosing these rotated controls is that they are symmetric with respect to the $xy$-plane, and hence the  FFs along $xy$ that they produce are equal $F_x(\omega,T)=F_y(\omega,T)$.
When symmetry of the $xy$ FFs is required, these controls can be conveniently used to achieve this feature. 

\subsection{Alternative IQ-control representation}
\label{sec:IQ_control}
The widespread use of the IQ-control suggests that, in some cases, it will be convenient to perform the optimization in this space. The polar IQ-control coordinate representation $\vOmega(t)=\Omega(t)(\cos\phi(t),\sin\phi(t),0)$, captures the degrees of freedom of the controls in the amplitude $\Omega(t)$ and phase $\phi(t)$ parameters.
The first step towards adapting F-GRAFS to work in the IQ-controls framework is to expand $\Omega(t),\phi(t)$ in terms of Slepians
\eqs{
\Omega(t) = \sum_k \alpha_k^\Omega v^{(k)}(t), \\
\phi(t) = \sum_k \alpha_k^\phi v^{(k)}(t).
}
Let us consider the objective function Eq.~(\ref{eq:objective_function}), and proceed as in Sec.~\ref{sec:FFF}.
The calculation is analogous, where Eq.~(\ref{eq:partials_Uc}) now turns into
\eqs{
\frac{\partial U_C(t)}{\partial \alpha_k^{\Omega}} = \sum_{n=0}^{N-1}\frac{\partial U_C(t)}{\partial \Omega_n} v_n^{(k)}, \\
\frac{\partial U_C(t)}{\partial \alpha_k^{\phi}} = \sum_{n=0}^{N-1}\frac{\partial U_C(t)}{\partial \phi_n} v_n^{(k)}.
}
The derivatives with respect to the IQ-control parameters now become, using the chain rule,
\eqs{
\frac{\partial U_C(t)}{\partial\Omega_n} &= \frac{\partial \Omega_n^x}{\partial\Omega_n}\frac{\partial U_C(t)}{\partial\Omega_n^x}+\frac{\partial\Omega_n^y}{\partial\Omega_n}\frac{\partial U_C(t)}{\partial\Omega_n^y} = \\
&=\cos\phi_n\frac{\partial U_C(t)}{\partial\Omega_n^x}+\sin\phi_n\frac{\partial U_C(t)}{\partial\Omega_n^y} ,\\
\frac{\partial U_C(t)}{\partial\phi_n} &= \frac{\partial \Omega_n^x}{\partial\phi_n}\frac{\partial U_C(t)}{\partial\Omega_n^x}+\frac{\partial\Omega_n^y}{\partial\phi_n}\frac{\partial U_C(t)}{\partial\Omega_n^y} = \\
&=-\Omega_n\sin\phi_n\frac{\partial U_C(t)}{\partial\Omega_n^x}+\Omega_n\cos\phi_n\frac{\partial U_C(t)}{\partial\Omega_n^y},
}
where $\Omega_n^x=\Omega_n\cos\phi_n$ and $\Omega_n^x=\Omega_n\sin\phi_n$.
The derivatives of $U_C(t)$ with respect to $\Omega_n^x,\Omega_n^y$ can be computed following Eq.~(\ref{eq:exp-partial}).

Numerical investigations using F-GRAFS were performed, comparing the IQ-control optimization scheme with the Cartesian one described in the main text. Results show that utilizing IQ-control in F-GRAFS yields solutions that achieve equivalent levels of cancellation in the CNB, when the controls are initialized as described in Appendix~\ref{sec:initial_conditions}.

\section{Filtering Single-axis Noise with Multi-axis Control F-GRAFS}
\label{sec:MC_SN}

In the main text, we analyze the single-axis noise optimization problem using both single- and multi-axis control F-GRAFS. 
The single-axis control (see Sec.~\ref{subsubsec:SA-optW}) presents a sharp improvement in optimization performance at a critical bandwidth of $W_c=2\times|\CNB| \times \delta t/2\pi$.
In the multi-axis control and multi-axis noise scenarios (see Sec.~\ref{subsubsec:ma-optw}), a sharp improvement in performance can be observed at a smaller control bandwidth, namely $1.5\times|\CNB|\times \delta t/2\pi$.
The single-axis noise configuration serves as a probe to analyze the advantages of multi-axis control, since it can be analyzed effectively with both control schemes. Since the multi-axis initial conditions described in Sec.~\ref{sec:MA_IC} reduce to CD when there is no control noise, CD initial conditions were used along the $x$ axis, while the $y$ axis was initialized with zero control amplitude.

In the present section, we study the optimized control powers compared to the initial conditions, and find no distinguishing increase in power at the critical bandwidth. Hence, we propose that the improvement in critical control bandwidth is due to the increase in control capabilities. Additionally, we find that the critical bandwidth is gate-dependent, and argue that CD is not the optimal initial condition for the single-axis noise and multi-axis control configuration.

\subsection{Optimized Power Analysis}
\label{subsec:MC_SN_power}

In principle, the sharp improvement of noise cancellation performance at the critical bandwidths could be given by a significant change in control amplitude. This could allow the control waveforms to reach different, possibly better, solutions in the objective function landscape. As we show below, no significant change in power is observed, suggesting that no additional complexity is gained by increasing the control amplitude.

\begin{figure}[h]
\centering
\includegraphics[width=.45\textwidth]{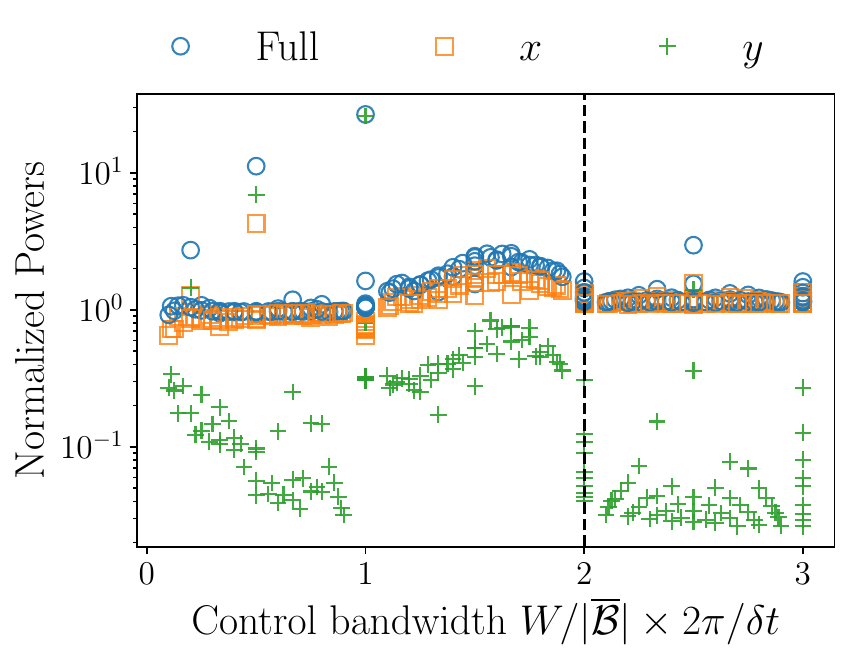}
\caption{Average power for single-axis noise with high-frequency cutoff $\omega_z$, solved using multi-axis control F-GRAFS.
Each point represents a different combination of $\omega_z$ and $W$, averaged over the Clifford+$T$ gate set, generating high-pass filters.
The initial conditions used correspond to CD along the $x$ direction and zero control along $y$.
The figure displays the full optimized control power (blue, circles)) as well as directional powers along $x$ (orange, squares) and $y$ (green, crosses), normalized by the initial condition power corresponding to each noise configuration. 
}
\label{fig:power_MC_MN}
\end{figure}

Fig.~\ref{fig:power_MC_MN} presents the optimized control powers of different noise and control configurations, averaged over the Clifford+$T$ gate set as function of $W$.
The directional control powers along $\sigma_x$ and $\sigma_y$ are defined as $P_{i}=\int_0^T \Omega_i(t)^2 dt$ for $i=x,y$ and the full control power as $P=\sum_{i=x,y}P_i$.
Optimizations are initialized with CD as $\Omega_{CD}(t)=\omega_z$ and the power values shown in the figure are normalized by its power $P_{CD}=\omega_z^2 T$.
Note that the $y$-axis controls (green crosses) need to be activated in order to produce general single-qubit rotations, but for $W>2|\CNB|\delta t/2\pi$, the amplitudes along $\sigma_x$ are larger by over an order of magnitude.
For $W>2|\CNB|\delta t/2\pi$, the control powers deviate little from the initial condition values (normalized power close to 1). This means that there is no significant change in control power between low and high bandwidth solutions. Since the initial conditions are maintained the same, this suggests that the improvement in performance is due entirely to the increased control capabilities given by the bandwidth.

\subsection{Optimization Dependence on Ideal Gate Fidelity Constraints}
\label{subsec:MC_SN_tolerances}

Another factor influencing the optimization performance is the tolerance $\tau_G$ set for the ideal gate fidelity, i.e., $\mF_G>1-\tau_G$.
Throughout the work described in the main text, $\tau_G$ was set to $10^{-10}$.
In this section, we show how in some cases relaxing this requirement leads to improvements in optimal bandwidth conditions.

\begin{figure}[h]
\centering
\includegraphics[width=.45\textwidth]{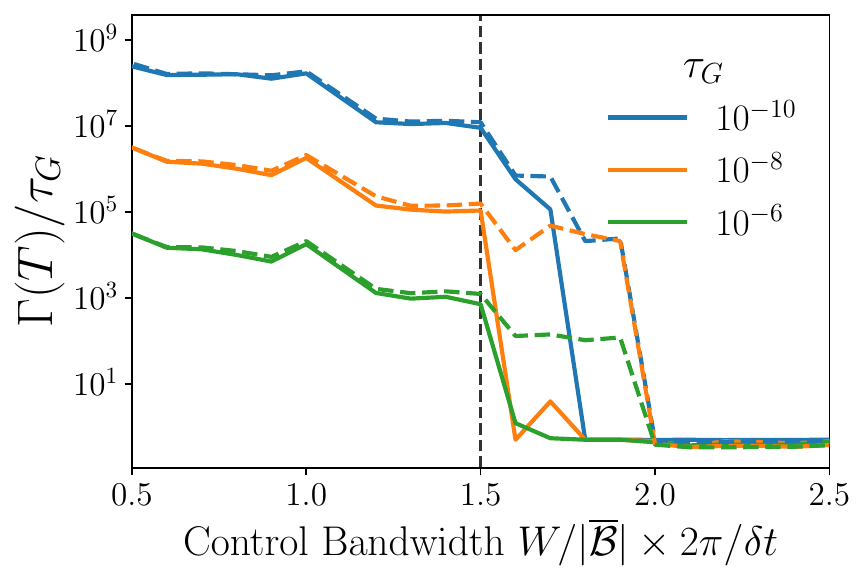}
\caption{
Dependence on tolerance for ideal gate fidelity $\tau_G$ of F-GRAFS performance, for the single-axis noise along $z$ case with multi-axis control along $x,y$.
Solid (dashed) lines represent the means (medians) over the Clifford+$T$ gate set.
In each case, data shown represent averages over bandwidth windows of size 0.1.
Optimized objective function values are normalized by $\tau_G$.
Note that after $W_c$ these values converge to 1, implying that the noise if filtered effectively and most of the infidelity is given by the gate constraints.
}
\label{fig:tolerance_MC_MN}
\end{figure}

Fig.~\ref{fig:tolerance_MC_MN} presents the bandwidth dependence of F-GRAFS performance when optimizing single-axis noise with multi-axis control.
The figure illustrates how the critical bandwidth $W_c$ changes with the tolerance for the ideal gate fidelity.
For all studied tolerances $\tau_G=10^{-10},10^{-8},10^{-6}$, the means over the Clifford+$T$ gate set (dashed lines) satisfy a $W_c=2\times\omega_z \times \delta t/2\pi$. 
The medians (solid lines) on the other hand, present a lower critical bandwidth, notably $W_c=1.5\times \omega_z \times \delta t/2\pi$ for $\tau_G=10^{-8},10^{-6}$, like in the multi-axis noise with multi-axis control case.
This difference between mean and median behavior implies that there exist numerous gates for which a bandwidth of $W>1.5\times \omega_z \times \delta t/2\pi$ is sufficient to achieve the desired noise cancellation.

The fact that this trade-off between gate fidelity and bandwidth constraints is gate-dependent, suggests that the CD initial conditions are not universally optimal for all gates.
In the presence of low-bandwidth constraints e.b. given by hardware, more gate-aware initial conditions are necessary in order to perform high degree of noise cancellation. 
To summarize, we observe that a degree of improvement in optimal bandwidth conditions can be obtained, albeit gate dependent, by relaxing the tolerance constraints on ideal gate fidelity.

\section{Effects of Noise PSD Specifications on F-GRAFS Optimized Controls}
\label{sec:PSD_comp}

In Sec.~\ref{sec:F-GRAFS} we describe the general procedure for finding a CNB from a given PSD, which involves the choice of a noise fractional power [Eq.~(\ref{eq:fractional-power})].
Intuitively, $\epsilon$ is the spectral leakage in the NB, meaning that $1-\epsilon$ corresponds to the amount of power that F-GRAFS-optimized controls will filter in the CNB.
Consequently, there is strong interest in choosing $\epsilon\ll1$ as small as possible.
In this section, we address the effect of $\epsilon$ on the optimization performance. More specifically, we investigate the effect that different choices of $\epsilon$ have on the operational fidelity given in Eq.~(\ref{eq:op-fidelity}). 
We show analytically and numerically that the loss in fidelity is to first order proportional to $\epsilon$.

\begin{figure}[h]
\centering
\includegraphics[width=.45\textwidth]{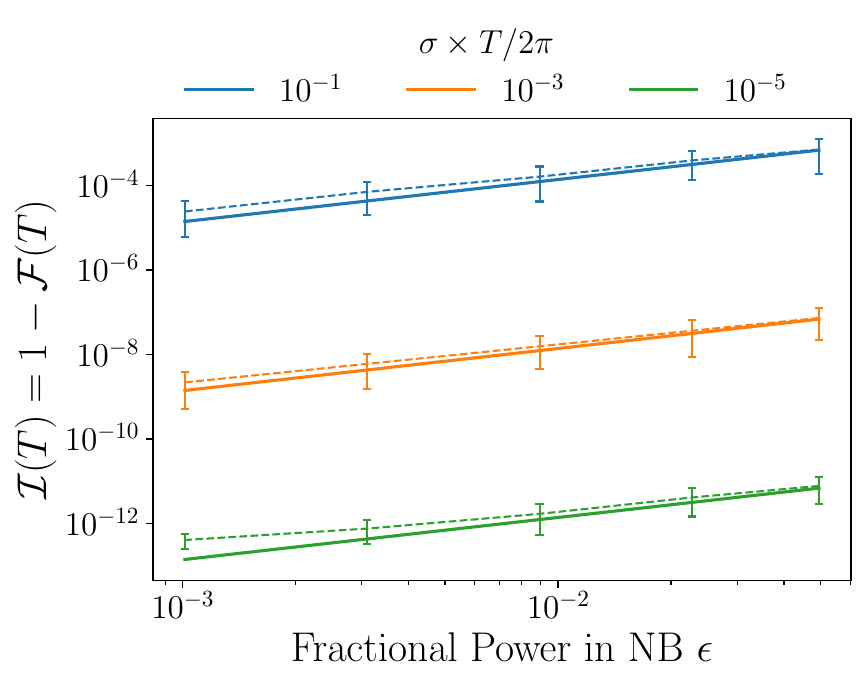}
\caption{(Dashed lines) Operational infidelities $\mathcal{I}(T)=1-\mathcal{F}(T)$ obtained from averaging over 1000 noise realizations of single-axis noise and control simulations. Errorbars represent standard deviations of these noise realizations.
Since the total noise power is equal to $2\pi\sigma^2$, varying the $\sigma$ is equivalent to changing the noise strength.
The fractional power in the CNB $\epsilon$ is varied through modifying the simulated noise correlation length $\gamma$, via the relationship $\gamma=\omega_H\tan(\epsilon\pi/2)$. The FF is kept constant with $\omega_H=0.01\times2\pi/\delta t$.
(Solid lines) Analytical predictions described in the text, see Eq.~(\ref{eq:analytical-infidelity-CNB}).}
\label{fig:PSD_CNB}
\end{figure}

In Fig.~\ref{fig:PSD_CNB} we show the values of infidelity $\mathcal{I}(T)=1-\mF(T)$ (dashed lines) for a qubit with single axis noise along $z$ and control along $x$. The average fidelity $\mF(T)$
is obtained from simulation, averaging over 1000 different noise realizations as well as two different gates: the identity and $X$ gates. 
The qubit is subject to F-GRAFS optimized controls and dephasing noise characterized by a Lorentzian spectrum with standard deviation $\sigma$ and correlation time $\tau\sim1/\gamma$.
The error bars represent the standard deviations of these different processes.

Throughout this analysis, the  FF is kept constant with $\omega_H=0.01\times2\pi/\delta t$ and tolerance of $10^{-15}$. 
The fractional power $\epsilon$ was varied in the range $[10^{-3},10^{-1}]$, which modified the degree to which the noise affects the qubit. 
In order to change the fractional power, the noise was modified by adapting $\gamma$ through the relationship $\gamma=\omega_H\tan(\epsilon\pi/2)$ (see Sec.~\ref{sec:simulation} for further details).

The solid lines in Fig.~\ref{fig:PSD_CNB} represent an analytical estimation of the values of the fidelity. This was obtained through approximating the FF as a delta function centered around $\omega_H$, namely $F(\omega)\approx \frac{T}{2} \delta(\omega-\omega_H)$. 
Using the explicit formula for the Lorenzian PSD, we compute the overlap
\eq{
\chi(T) &= \frac{1}{\pi}\int_0^\infty S(\omega) F(\omega) d\omega \\
&\approx \frac{1}{\pi} \int_0^\infty \frac{2\gamma\sigma^2}{\gamma^2+\omega^2} \times \frac{T}{2} \delta(\omega - \omega_H) d\omega = \\
&= \frac{T \sigma^2}{\pi} \frac{\gamma }{\gamma^2+\omega_H^2}.
}
Using the relationship between $\gamma$ and $\epsilon$, we rewrite this expression in terms of $\epsilon$ instead of $\gamma$ for fixed $\omega_H$:
\eq{
\chi(T) &= \frac{T\sigma^2 }{\pi} \frac{\sin(\pi\epsilon)}{2\omega_H} \\
&\approx  \frac{T \sigma^2}{2 \omega_H} \epsilon,
}
where the last line holds when $\epsilon\ll\pi$.
This allows us to estimate the fidelity as a function of the fractional power $\epsilon$. In the case of high gate fidelity (in this case set to $10^{-15}$), the infidelity will be
\eq{
\label{eq:analytical-infidelity-CNB}
\mathcal{I}(T) &= 1 - \mF(T) = 1- \left\langle \Big| \frac{1}{2} \tr U_G^\dagger U(T) \Big|^2 \right\rangle \\
& \approx 1-\left\langle \Big| \frac{1}{2} \tr \tU(T) \Big|^2 \right\rangle = 1-\frac{1+e^{-\chi(T)}}{2} \approx \frac{\chi(T)}{2} \\
&\approx \frac{T \sigma^2}{4 \omega_H} \epsilon.
}
This relationship is shown in Fig.~\ref{fig:PSD_CNB} as solid lines, with excellent agreement with the results of the simulations. This further supports the approximation of the F-GRAFS FFs obtained from CD initial conditions as delta functions centered around the frequency cutoff.

It is worth noting that the fidelity decays linearly with the factor $\sigma^2 \epsilon$, corresponding to the noise power in the CNB. This result suggests that although detailed knowledge of the PSD is not required for F-GRAFS optimization, minimizing the noise contributions to the CNB is crucial in order to maintain high fidelity gates.

\section{Alternative Objective Functions}
\label{sec:alternative_objective_functions}
The benefit of the objective function Eq.~(\ref{eq:objective_function}) is that it only requires limited knowledge of the noise PSD. 
This is, the only feature of the noise needed to compute the optimization is the location of the null band $\mathcal{B}$ and the CNB.
In this section, we adapt the algorithm to include two variations of F-GRAFS.
More specifically, when detailed knowledge of the noise PSD is at hand, the overlap itself can be used as an objective function.
Furthermore, we show via numerical experiments that it is possible to generate localized but targeted FFs, such as Gaussian-shaped FFs.
For simplicity, in what follows we will work in the single-axis noise and control scenario.

\subsection{Explicit Spectrum Filtering}
An alternative to the objective function Eq.~(\ref{eq:objective_function}) can be devised when detailed information of the PSD is available.
If $S(\omega)$ is known, then it is possible to define the following objective function
\eq{
\Gamma^{\mrm{PSD}}(T;S) = \int_{0}^\infty  S(\omega) F(\omega,T) d\omega.
}
Since the functional dependence on the control expansion parameters has not changed, the  computation of the gradients follows as in Sec.~\ref{sec:FFF}. It is straightforward to see that the derivative of the FF appears in the gradient in the following way
\eq{
\frac{\partial}{\partial \alpha_k} \Gamma^{\mrm{PSD}}(T;S) =
\int_{0}^\infty  S(\omega)\frac{\partial}{\partial \alpha_k} F(\omega,T) d\omega
}

In Fig.~\ref{fig:alternative-ofs} panels (a) and (b) we present results from this optimization, that takes into account the specifics of the PSD (PSD-F-GRAFS). We compare with the results obtained with the regular F-GRAFS optimization. 
On panel (b), the noise PSD (black), was defined as 
$S(\omega) = A/\omega^2$ with hard cutoff at a given value $\omega_H$,
where the constant $A$ is chosen to match the amplitude of the FF. This PSD also presents a low frequency cutoff, to prevent DC divergence.

From Fig.~\ref{fig:alternative-ofs}(b) it is clear that while both achieve greater cancellation than pure CD (green), the PSD-F-GRAFS version follows an inverse trend compared to the PSD.
With detailed information of the PSD, it is possible to get finer grained features in the FF that achieve greater cancellation over those frequencies where the noise is stronger. 
On the other hand, the F-GRAFS solution aims to get average cancellation over the entire band and is considerably more flat in the CNB.
Additionally, both optimizations achieve equivalent objective function values, $\Gamma^{\mrm{PSD}}(T)\sim P\times \Gamma^{\mrm{F-GRAFS}}(T,S) \sim 10^{-10}$, where $P$ is the total power of the noise $S(\omega)$. 
It took the F-GRAFS method about 25 steps to achieve this level of cancellation, while the PSD-F-GRAFS method only took 6 iterations.
This highlights that although finer resolution of the PSD can improve the computational cost, it is not necessary and F-GRAFS can find equally good solution for all the cases we studied. This is further substantiated in Fig.~\ref{fig:alternative-ofs}(a), where we can see that the control functions obtained through both methods are very similar. 

\subsection{Target Filter Functions}

Alternatively, one could ask whether FFs can be specifically shaped to a given targeted objective FF $F_{\mrm{target}}(\omega)$.
For this application, we define the objective function as the quadratic difference between the FF and the targeted function,
\eq{
\Gamma^{\mrm{target}}(T;F_{\mrm{target}}) = \int_{0}^\infty  (F(\omega,T) - F_{\mrm{target}}(\omega))^2 d\omega.
}
As in the previous section, the computation of the gradient proceeds in the same way outlined in Sec. \ref{sec:FFF}, where the derivative of the FF is obtained by
\eq{
&\frac{\partial}{\partial \alpha_k} \Gamma^{\mrm{target}}(T;F_{\mrm{target}}) = \\
&=2\int_{0}^\infty (F(\omega,T) - F_{\mrm{target}}(\omega)) \frac{\partial}{\partial \alpha_k} F(\omega,T)d\omega .
}
\begin{figure}[h]
\centering
\includegraphics[width=.48\textwidth]{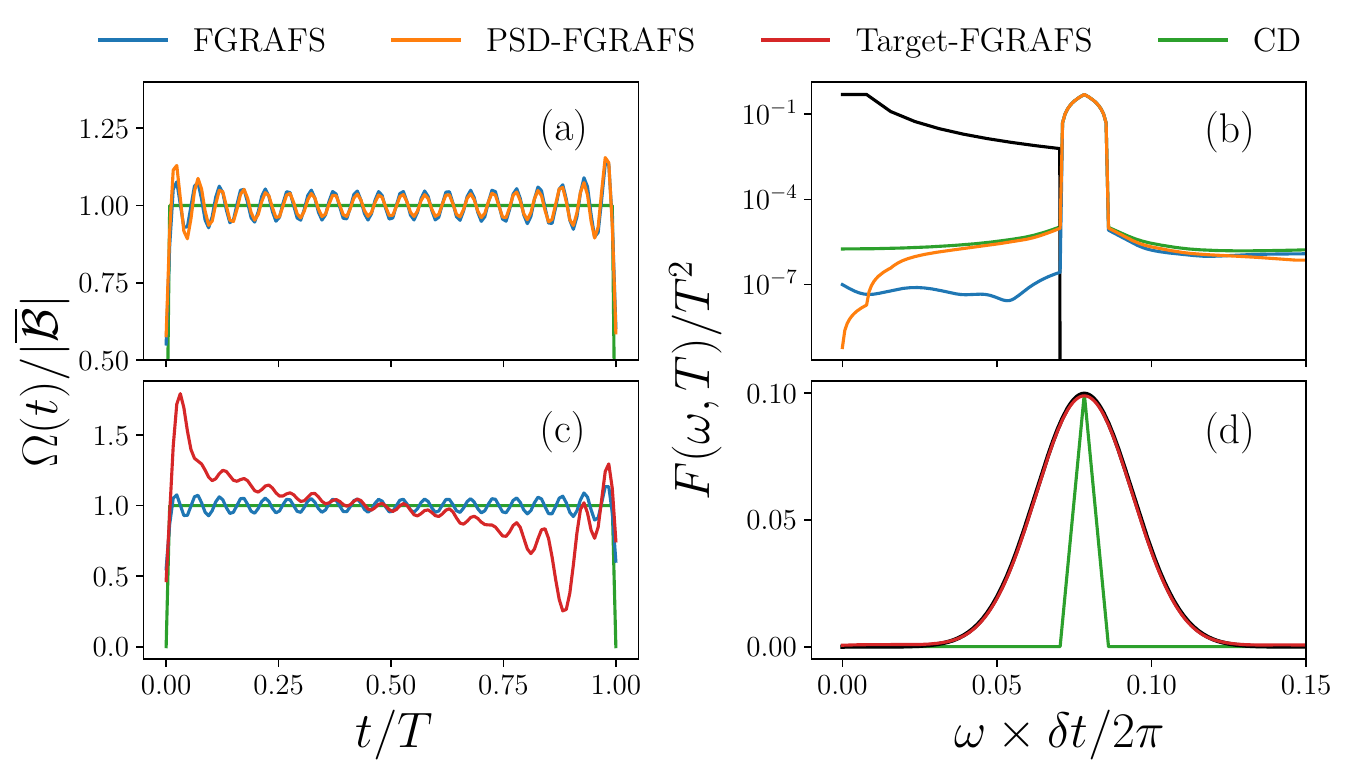}
\caption{
(Top row) Comparison between controls (a) and FFs (b) obtained from F-GRAFS and the PSD-F-GRAFS optimizations. 
The noise PSD, shown in black, is defined as $1/f$-type noise with hard cutoff at $\omega_H=0.08\times2\pi/\delta t$, where $S(\omega) = A/\omega^2$, where the constant $A$ is chosen to match the amplitude of the filter function. The PSD $S(\omega)$ presents an additional cutoff at low frequencies.
Both the F-GRAFS (blue) and the PSD-F-GRAFS (orange) achieve greater cancellation than the initial condition with CD (green). The PSD-F-GRAFS FF presents more features in frequency, due to the access to higher PSD resolution. 
(Bottom row) Optimization results of controls (a) and FFs (b) obtained with Target-F-GRAFS (red), with target function $F_{\mrm{target}}(\omega)=A\exp(-(\omega-\omega_0)^2/2\sigma^2)$. 
In (b), it is clear that the target-F-GRAFS FF (red) approximates well the target function (black). In green we again show the initial condition resulting from CD control.}
\label{fig:alternative-ofs}
\end{figure}
In Fig.~\ref{fig:alternative-ofs} panels (c) and (d) we present an example of this optimization for a gaussian target function $F_{\mrm{target}}(\omega)=A\exp(-(\omega-\omega_H)^2/2\sigma^2)$. 
The F-GRAFS initial conditions were set such that the FF is centered around the same value $\omega_H$ where the target Gaussian is located.
In panel (d), the black curve represents the target Gaussian function. We can see that F-GRAFS is capable of reshaping the initial constant drive condition into a Gaussian function with good agreement. 
Note that the target functions need to be normalized such that the total area of the FF is the total time $T$, which means that $A=T/\sigma \sqrt{2\pi}$.

\bibliographystyle{apsrev4-1}
%
\end{document}